%% file: main.tex
\documentclass[lettersize,journal]{IEEEtran}
\usepackage{hyperref}
\usepackage{amsmath,amsfonts}
\usepackage{array}
\usepackage{textcomp}
\usepackage{stfloats}
\usepackage{url}
\usepackage{verbatim}
\usepackage{graphicx}
\usepackage{cite}
\usepackage{amsmath}
\usepackage{amsthm}
\usepackage{pifont}
\usepackage{diagbox} 
\usepackage{array}   
\usepackage{multirow}
\usepackage{booktabs}
\usepackage{enumitem} 
\usepackage{tikz}
\usepackage{subcaption}
\usepackage{float}
\usepackage{xparse}
\usepackage{mathtools}
\usepackage[ruled, noend]{algorithm2e}
\usepackage{makecell}
\usepackage{siunitx}  
\newtheorem{assumption}{Assumption}
\newtheorem{theorem}{Theorem}

\NewDocumentEnvironment{pot}{m}{%
  \noindent\textbf{Proof of Theorem~\ref{#1}.} 
}{%
  \qed 
}

\begin{document}

\title{DistJoin: A Decoupled Join Cardinality Estimator based on Adaptive Neural Predicate Modulation}

%\author{IEEE Publication Technology,~\IEEEmembership{Staff,~IEEE,}
\author{
\IEEEauthorblockN{
Kaixin Zhang,
Hongzhi Wang\textsuperscript{*}\thanks{Corresponding author},
Ziqi Li,
Yabin Lu,
Yu Yan,
Yingze Li,
Yiming Guan
}
\IEEEauthorblockA{
\\Harbin Institute of Technology\\
Email: 21B903037@stu.hit.edu.cn, wangzh@hit.edu.cn, {24B903030,22S003043}@stu.hit.edu.cn, yuyan@hit.edu.cn,{23B903046,24S103297}@stu.hit.edu.cn
}
}
        % <-this % stops a space
%\thanks{This work is supported by National Natural Science Foundation of China (NSFC) (62232005).}% <-this % stops a space
%\thanks{Manuscript received April 19, 2021; revised August 16, 2021.}}

\markboth{Journal of \LaTeX\ Class Files,~Vol.~14, No.~8, August~2021}%
{Shell \MakeLowercase{\textit{et al.}}: A Sample Article Using IEEEtran.cls for IEEE Journals}


\maketitle
%\footnotetext[2]{} 
\begin{abstract}
    Research on learned cardinality estimation has made significant progress in recent years. However, existing methods still face distinct challenges that hinder their practical deployment in production environments. We define these challenges as the ``Trilemma of Cardinality Estimation'', where learned cardinality estimation methods struggle to balance generality, accuracy, and updatability. To address these challenges, we introduce DistJoin, a join cardinality estimator based on efficient distribution prediction using multi-autoregressive models. Our contributions are threefold: (1) We propose a method to estimate join cardinality by leveraging the probability distributions of individual tables in a decoupled manner. (2) To meet the requirements of efficiency for DistJoin, we develop Adaptive Neural Predicate Modulation (ANPM), a high-throughput distribution estimation model. (3) We demonstrate that an existing similar approach suffers from variance accumulation issues by formal variance analysis. To mitigate this problem, DistJoin employs a selectivity-based approach to infer join cardinality, effectively reducing variance. In summary, DistJoin not only represents the first data-driven method to support both equi and non-equi joins simultaneously but also demonstrates superior accuracy while enabling fast and flexible updates. The experimental results demonstrate that DistJoin achieves the highest accuracy, robustness to data updates, generality, and comparable update and inference speed relative to existing methods.
\end{abstract}


\section{Introduction}
\label{section:Introduction}
\IEEEPARstart{C}{ardinality} estimation is a critical task in DBMSs, as its precision directly affects downstream tasks such as cost prediction~\cite{QPPNet, QueryFormer, ZeroShot, MB2} and query optimization~\cite{NEO, BAO}. Researchers have proposed numerous conventional histogram-based and sampling-based methods~\cite{TradCE1, SortingCE, BloomFilterCE, SamplingCE1, SamplingCE2, SamplingCE4, SamplingCE5, SamplingCE6, HistogramCE1, HistogramCE2, HistogramCE3, HistogramCE4, HistogramCE5, HistogramCE6, HistogramCE7, HistogramCE8}. However, due to the inherent complexity of the problem, conventional cardinality estimators have long suffered from significant estimation errors~\cite{ExpCESuvery, AreWeReady}.

In recent years, with the advancement of deep learning, numerous learned cardinality estimators~\cite{AreWeReady} have been proposed, achieving substantially higher accuracy than conventional approaches. These learned estimators can be broadly classified into three types: query-driven, data-driven, and hybrid methods. 

Query-driven methods~\cite{LWXGB, MSCN, RobustMSCN, ALECE, QueryDriven2, QueryDriven3, QueryDriven4, QueryDriven5} leverage workload data for flexibility, but face accuracy challenges and updatability limitations due to workload dependency. While some works~\cite{RobustMSCN, ALECE} mitigate updatability by incorporating data features, they still require costly workload generation. Last but not least, the accuracy of query-driven methods is limited, especially when the true cardinality of queries lies in an extremely wide range, such as in non-equal join query scenarios.

Data-driven methods~\cite{DeepDB, FLAT, Naru, NeuroCard, Duet, FACE, ASM, BayesCard, FactorJoin} learn from data, enhancing robustness and accuracy, but face technical-path-specific weaknesses. SPN-based methods~\cite{DeepDB, FLAT} and autoregressive models~\cite{Naru, NeuroCard, Duet} struggle with non-equi joins and update flexibility due to full outer join dependencies. For Instance, NeuroCard~\cite{NeuroCard} requires full retraining for single-table updates. FactorJoin~\cite{FactorJoin} improves flexibility by decoupling the model, but sacrifices accuracy and support for non-equi joins due to its binning and join cardinality inference.

Hybrid methods~\cite{UAE, Duet} typically build upon data-driven approaches while incorporating both supervised and unsupervised learning techniques to further enhance estimation precision. However, Duet mainly focuses on single-table queries, while UAE's training time and memory overhead are huge due to its hybrid training process.

In summary, as shown in \autoref{tab:methods_comparsion}, previous methods cannot simultaneously support three key features: generality, accuracy, and updatability. The fundamental limitation of query-driven methods lies in their reliance on supervised learning from workloads, which requires costly dataset collection and struggles with the i.i.d. assumption~\cite{Duet}. The hybrid methods obtain higher accuracy by employing hybrid training, which significantly increases their training costs and does not efficiently support join queries. 

Given the inherent challenges of query-driven and hybrid approaches, our work focuses on data-driven approaches. Autoregressive models have been shown to outperform SPNs in single-table distribution estimation~\cite{Naru, NeuroCard}. However, when directly applied to multiple tables, they face the Full Outer Join dependency issue. Therefore, we employ a decoupled design to preserve their advantages while circumventing their limitations.

To accomplish this objective, we present DistJoin\footnote{Source code available here: \url{https://github.com/GIS-PuppetMaster/DistJoin}}, an innovative join cardinality estimator built on decoupled multi-autoregressive modeling. Intuitively, DistJoin first estimates the distributions of each table's join key column's distinct values. Then, instead of calculating each join key's count like FactorJoin~\cite{FactorJoin}, DistJoin calculates each key's normalized probability to infer their weights in the join schema (a.k.a. corresponding join query without any predicates) with a weighted-sum formula. Finally, DistJoin computes the join schema's true cardinality using maintained join key distributions and combines this with the normalized weights via inner product to derive the final estimate.

DistJoin's decoupled design fundamentally advances prior methods by simultaneously achieving generality, accuracy, and updatability-a combination previously unattained. Unlike FactorJoin's reliance on full-outer joins and binning, DistJoin's selectivity-based inference operates directly on join key distributions, enabling support for both equi and non-equi joins while minimizing error propagation through probability-weighted aggregation. Additionally, its per-table autoregressive models permit independent updates, addressing a critical limitation of monolithic approaches like NeuroCard. To fully realize these advantages, however, two interconnected challenges must be addressed: precise and scalable join cardinality inference, along with efficient distribution computation for high-dimensional join keys.

\begin{enumerate}[label=\alph*)]
    \item \textbf{Precise and scalable join cardinality inference.} Previous methods like FactorJoin compute join cardinality directly from binned join key counts of individual tables under given predicates. We term this process of deriving multi-table join cardinality from single-table estimations as \textit{join cardinality inference}. The binning process seriously damages its accuracy and prevents it from supporting non-equi join. Besides, in \autoref{subsection:FactorJoinVarAnalysis}, we demonstrate how this approach leads to variance accumulation from individual table estimators, ultimately compromising accuracy and scalability for large, complex join queries.

    \item \textbf{Achieving highly efficient distribution estimation.} DistJoin requires efficient estimation of join key distributions under arbitrary predicates to support non-equi join queries, flexible updates, and accurate cardinality inference. Although our previous work Duet~\cite{Duet} solved NeuroCard's inference complexity problem~\cite{NeuroCard,Duet} caused by its progressive sampling, the framework of Duet is not competitive with NeuroCard's lossless column factorization technique, leading to low efficiency in large NDV(number of distinct values) situations.
\end{enumerate}

To address the first challenge, DistJoin employs a \textit{selectivity-based join cardinality inference} process that operates on estimated complete join key distributions under given predicates. This distribution-based approach offers four key advantages:
\begin{enumerate}[label=\alph*)]
    \item The method naturally supports diverse join conditions, including equi joins, non-equi joins, and outer joins.
    \item The join cardinality inference process introduces no additional error sources beyond single-table distribution prediction errors and their corresponding accumulation, unlike techniques such as binning. The variance accumulation problem is also significantly alleviated. Furthermore, it operates independently of any specific assumptions, unlike most conventional methods.
    \item The decoupled method relies solely on individual table distribution predictions from independently updatable models.    
\end{enumerate}

To address the last challenge, we propose the Adaptive Neural Predicate Modulation (ANPM) model, extending our previous work in Duet~\cite{Duet}. ANPM employs HyperNetworks and low-rank techniques to generate MLP models that modulate sub-column information into the logits predicted by autoregressive models. The ANPM model addresses three critical aspects to provide support for DistJoin's decoupled join cardinality estimation:
\begin{enumerate}[label=\alph*)]
    \item Adapting the lossless column factorization technique~\cite{NeuroCard} to Duet's framework.
    \item Effectively reconstructing original column distributions.
    \item Addressing MADE model limitations~\cite{MADE, Naru}, specifically: (1) insufficient fitting capability for the first column due to its input-independent nature, and (2) difficulty in adapting to skewed distributions.
\end{enumerate}

Our primary contributions include:
\begin{enumerate}[label=\alph*)]
    \item We present DistJoin, a novel join query cardinality estimator based on multiple decoupled estimators of individual table distributions. To our knowledge, DistJoin is the first data-driven method capable of handling both equi and non-equi join queries, which benefits from independent join key distribution prediction and low-variance selectivity-based join cardinality inference.

    \item We conduct a comprehensive error analysis comparing FactorJoin's and DistJoin's join cardinality inference approaches. Our results validate FactorJoin's variance accumulation issue and demonstrate that DistJoin's join cardinality inference effectively mitigates this problem, as its variance increases more slowly than FactorJoin's approach.

    \item To ensure computational efficiency in training and inference, we introduce ANPM, which utilizes HyperNetworks and low-rank decomposition techniques to generate MLP models for sub-column information modulation. ANPM enables efficient and accurate estimation of join key distributions under predicates while effectively scaling across tables with varying column counts and domain sizes. This approach eliminates the accuracy and generality trade-offs inherent in existing methods.

    \item Extensive experimental evaluation demonstrates DistJoin's superiority over common baselines in accuracy, robustness, and generality while maintaining competitive training speed.
\end{enumerate}

In the rest of this paper, we first detail analyze the background of previous methods and clarify the motivation and preliminaries in \autoref{section:Background}. We then present an overview of DistJoin's design in \autoref{section:Overview}, followed by a comprehensive error analysis and design of DistJoin's join cardinality inference approach in \autoref{section:joincard_est}. Then, we explain the design of the ANPM model, which predicts the distribution needed by the join cardinality inference in \autoref{section:dist_estimation}. Finally, we evaluate DistJoin's performance against common baselines in \autoref{section:Experiment} and conclude the paper in \autoref{section:Conclusion}.

\section{Motivation \& Preliminaries}
\label{section:Background}

\subsection{Background \& Motivation}

\begin{table*}
\footnotesize
\caption{Common methods comparison, `--' represents that a part of the corresponding methods can support the feature. There is a lack of a method that can support all three features. We discuss this in detail in ~\autoref{section:Background} for each method.}
\label{tab:methods_comparsion}
\centering
\begin{tabular}{|ll|cc|ccc|c|}
\hline
\multicolumn{2}{|l|}{\multirow{2}{*}{\diagbox[width=6cm, height=0.65cm, innerleftsep=1cm, innerrightsep=1cm]{\vspace{0cm}\normalsize Features}{\vspace{-0.6cm}\normalsize Methods}}}& \multicolumn{2}{c|}{Conventional Methods}                                                                                         & \multicolumn{3}{c|}{Learned Methods}                                                                                                                                                                                                                                                          & \multirow{2}{*}{\begin{tabular}[c]{@{}l@{}}DistJoin\\ (Ours)\end{tabular}} \\ \cline{3-7}
\multicolumn{2}{|l|}{}                                                                                                                                                              & \multicolumn{1}{c|}{\begin{tabular}[c]{@{}l@{}}Histogram\end{tabular}}   & \begin{tabular}[c]{@{}l@{}}Sampling\end{tabular}       & \multicolumn{1}{c|}{\begin{tabular}[c]{@{}l@{}}Query-Driven\end{tabular}} & \multicolumn{1}{l|}{\begin{tabular}[c]{@{}l@{}}Data-Driven\end{tabular}} & \begin{tabular}[c]{@{}l@{}}Hybrid\end{tabular}  &                 \\ \hline
\multicolumn{1}{|l|}{\multirow{2}{*}{Generality}}                                                                                      & Support on Equi Joins                      & \multicolumn{1}{c|}{\ding{51}}                                           & \ding{51}                                              & \multicolumn{1}{c|}{\ding{51}}                                            & \multicolumn{1}{c|}{\ding{51}}                                           &--                                               & \ding{51}       \\ \cline{2-8} 
\multicolumn{1}{|l|}{}                                                                                                                 & Support on Non-Equi Joins                  & \multicolumn{1}{c|}{\ding{51}}                                           & \ding{51}                                              & \multicolumn{1}{c|}{--}                                                   & \multicolumn{1}{c|}{\ding{55}}                                           &\ding{55}                                        & \ding{51}       \\ \hline
\multicolumn{1}{|l|}{\multirow{3}{*}{Accuracy}}                                                                                        & Relative Low Estimation Error              & \multicolumn{1}{c|}{\ding{55}}                                           & \ding{55}                                              & \multicolumn{1}{c|}{--}                                                   & \multicolumn{1}{c|}{--}                                                  &\ding{51}                                        & \ding{51}       \\ \cline{2-8} 
\multicolumn{1}{|l|}{}                                                                                                                 & Scalability with Cardinality               & \multicolumn{1}{c|}{\ding{51}}                                           & --                                                     & \multicolumn{1}{c|}{\ding{55}}                                            & \multicolumn{1}{c|}{--}                                                  &\ding{55}                                        & \ding{51}       \\ \cline{2-8} 
\multicolumn{1}{|l|}{}                                                                                                                 & Scalability with \#Joins                   & \multicolumn{1}{c|}{\ding{55}}                                           & --                                                     & \multicolumn{1}{c|}{\ding{51}}                                            & \multicolumn{1}{c|}{--}                                                  &\ding{55}                                        & \ding{51}       \\ \hline
\multicolumn{1}{|l|}{\multirow{3}{*}{Updatability}}                                                                                    & Adapt to Workload Drift                    & \multicolumn{1}{c|}{\ding{51}}                                           & \ding{51}                                              & \multicolumn{1}{c|}{\ding{55}}                                            & \multicolumn{1}{c|}{\ding{51}}                                           &\ding{51}                                        & \ding{51}       \\ \cline{2-8} 
\multicolumn{1}{|l|}{}                                                                                                                 & Fast Training and Update                   & \multicolumn{1}{c|}{\ding{51}}                                           & \ding{51}                                              & \multicolumn{1}{c|}{\ding{55}}                                            & \multicolumn{1}{c|}{--}                                                  &\ding{55}                                               & \ding{51}       \\ \cline{2-8}                                                 
\multicolumn{1}{|l|}{}                                                                                                                 & Flexible Update                            & \multicolumn{1}{c|}{\ding{51}}                                           & \ding{51}                                              & \multicolumn{1}{c|}{\ding{55}}                                            & \multicolumn{1}{c|}{--}                                                  &\ding{55}                                               & \ding{51}       \\ \hline
\end{tabular}
\vspace{-20pt}
\end{table*}

As we discussed in \autoref{section:Introduction}, several unresolved issues continue to hinder the practical deployment of learned cardinality estimators, primarily concerning generality, accuracy, and updatability. As illustrated in \autoref{tab:methods_comparsion}, existing methods can address one or two of these features, but often at the expense of the others. We conceptualize this challenge as the ``Trilemma of Cardinality Estimation" and analyze existing methods' support for its three aspects:

(\romannumeral 1) Generality. Generality refers to the range of query templates supported by a cardinality estimation method. While recent advancements in learned estimators have extended support to join queries, existing methods primarily focus on equi joins and struggle to support non-equi join queries effectively, especially for data-driven methods.

Data-driven methods based on SPNs~\cite{DeepDB, FLAT} rely entirely on full outer join table fanout values to capture inter-table correlations. As full outer joins are inherently equi joins, these methods cannot handle non-equi joins. NeuroCard~\cite{NeuroCard} supports join queries by learning from a comprehensive full outer join view, training on sampled tuples that include fanout information. While this enables equi-join schema predictions, NeuroCard similarly cannot support non-equi joins. FactorJoin~\cite{FactorJoin} employs a binning strategy with Bayesian Networks to estimate bin probabilities under given predicates for individual tables, using factor graphs for cardinality calculation. However, by ignoring intra-bin key distributions to maintain reasonable estimation times, FactorJoin cannot effectively support non-equi joins. Although query-driven methods technically support various join conditions, their accuracy significantly degrades with wide cardinality ranges due to numerical precision limitations, as demonstrated in \autoref{subsection:accuracy_evaluation}.Hybrid methods like UAE~\cite{UAE} doesn't support non-equi join queries due to it uses the same join estiamtion approach as NeruoCard.

(\romannumeral 2) Accuracy. Although learned estimators generally achieve higher accuracy than conventional methods, some approaches compromise accuracy to maintain efficiency or updatability. Beyond mean estimation error, the scalability of model accuracy is crucial, particularly in addressing long-tail error distributions and managing error growth rates across varying join sizes and cardinality ranges.

Query-driven methods~\cite{MSCN,ALECE} typically exhibit lower accuracy than modern data-driven methods in scenarios with both equi and non-equi join queries. The extreme cardinality ranges (from 1 to 1e32, as shown in \autoref{tab:Workloads}) present substantial challenges. Even with logarithmic transformation and Min-Max normalization, such ranges demand exceptional model precision. Our experiments reveal that MSCN, as a classic query-driven method, produces unusable errors for non-equi joins. While the most representative query-driven method, ALECE, also suffers from a serious long-tail error distribution problem. Among data-driven methods, autoregressive approaches like NeuroCard~\cite{NeuroCard} generally outperform SPN-based methods but still face challenges with sampling coverage in large fully connected views~\cite{STATS_CEB} and invalid column value combinations. FactorJoin~\cite{FactorJoin} suffers from relatively high errors due to its binning strategy and its join cardinality inference design.

(\romannumeral 3) Updatability. Model updating remains a significant challenge for the practical deployment of learned cardinality estimators. An ideal estimator should demonstrate low training costs and support flexible, on-demand updates while adapting to workload drift to minimize update frequency.

Query-driven methods, despite their generality, face significant update challenges due to their reliance on expensive query execution for training workload generation. Workload drift or data updates necessitate complete retraining with regenerated workloads. While data-driven methods avoid workload drift issues, they struggle to balance update flexibility with accuracy. SPN-based methods like DeepDB~\cite{DeepDB} and FLAT~\cite{FLAT} offer easy updates but lag in accuracy. Autoregressive methods like NeuroCard~\cite{NeuroCard} couple models with entire database full outer join views, requiring complete resampling and retraining even for single-table distribution changes. FactorJoin~\cite{FactorJoin} improves upon this by decoupling models from the full database, achieving greater update flexibility at the cost of accuracy. Hybrid methods like UAE inherit the training method of NeuroCard and therefore have the same problems.

\textbf{\underline{Motivation.}} As summarized in \autoref{tab:methods_comparsion}, existing methods struggle to balance generality, accuracy, and updatability. We also observe that data-driven approaches vary in their support for the aforementioned features. While no single method can simultaneously address all features, except non-equi joins, each feature is supported by at least one data-driven approach. Therefore, building upon advanced data-driven methods like FactorJoin, we aim to resolve its existing limitations to support all the above features simultaneously.

\subsection{Preliminary}
\label{subsection: Preliminary}

\textbf{\underline{AR-Based Cardinality Estimator.}} In cardinality estimation area, AR-based methods ~\cite{Naru, NeuroCard} employ autoregressive models to predict column value distributions under given predicates, denoted as $P(\mathcal{D}(C_i)|x_{<i})$, where $\mathcal{D}(C_i)$ represents the domain (distinct value set) of column $C_i$, and $x_{<i}$ denotes predicate values on columns preceding $C_i$ in the autoregressive order (e.g., ``3" in ``$C_i>3$"). These methods use progressive sampling for range predicates, which increases computational cost and error rates.

DistJoin's ANPM model is based on the framework of Duet~\cite{Duet}. Duet improves upon the above approach by incorporating complete predicates as input and estimating $P(\mathcal{D}(C_i)|pred_{<i})$ to eliminate sampling during inference. This enhancement significantly accelerates prediction, reduces GPU memory requirements, and mitigates error accumulation. Duet's selectivity estimation for query $\mathcal{Q}(T)$ is formally represented as:
\vspace{-5pt}
\begin{align}
\hat{p}(\mathcal{Q}(T)) &= \prod_{i=1}^{n} \sum_{v \in \mathcal{D}(T.C_i)} I(pred_i, v)\hat{p}_{T}(T.C_i=v|pred_{<i}) \label{eq:selection} \\
I(pred_i, v) &= \begin{cases} 
                    1 & \text{if $v$ satisfies $pred_i$} \\ 
                    0 & \text{otherwise}
                  \end{cases} \label{eq:indicator}
\end{align}
where $n$ denotes the number of columns in $T$.

Duet's training process involves randomly sampling tuples from training data and generating predicate sets that constrain these tuples as model input. The model is trained using cross-entropy loss targeting the sampled tuples.

\section{Overview}
\label{section:Overview}
DistJoin addresses the cardinality estimation problem for queries of the form:

``SELECT COUNT(*) FROM $T_1$, ..., $T_n$ WHERE $T_i.C_{key}$ $\diamond$ $T_j.C_{key}$ AND ... AND $\mathcal{Q}(T_k)$ AND ...;" 

Here, $T_i$ represents a relation, $\diamond \in \{=,>=,<=,>,<\}$ denotes the join condition, and $\mathcal{Q}(T_i)$ specifies a set predicates $\mathcal{P}$, and $\mathcal{Q} = \{\mathcal{Q}(T_i)|i=0,\dots,n\}$. Queries where all $\diamond$ operators are `$=$' are classified as equi joins, which most existing methods support. Conversely, queries containing other comparison operators represent non-equi joins, which current methods have yet to support effectively.

\underline{\textit{\textbf{Core Idea}}}: We approach join cardinality estimation from a probability distribution perspective. Capturing inter-table correlations is crucial to avoid errors from independence assumptions. To naturally represent these correlations, we estimate the join key distribution $\hat{P}(T_i.C_{key}, Q(T_i))$, where $C_{key}$ denotes the join key column. Despite the challenge of large domain sizes $|\mathcal{D}(T_i.C_{key})|$, we achieve precise and efficient estimation through our proposed ANPM.

For two-table equi joins, our \textit{selectivity-based join cardinality inference} operates as \autoref{equ:two_table_equi_join} and \autoref{equ:get_card}. \autoref{equ:two_table_equi_join} demonstrates the predicate selectivity calculation on joined results. We provide a detailed discussion in ~\autoref{section:joincard_est}. Intuitively, DistJoin repeats the process of ~\autoref{equ:two_table_equi_join} to implement a bottom-up join process simulation along the query plan tree, computing join selectivity in a vectorized, cache-efficient manner across layers.

\begin{figure}[!htbp]
    \centering
    \includegraphics[width=0.9\linewidth]{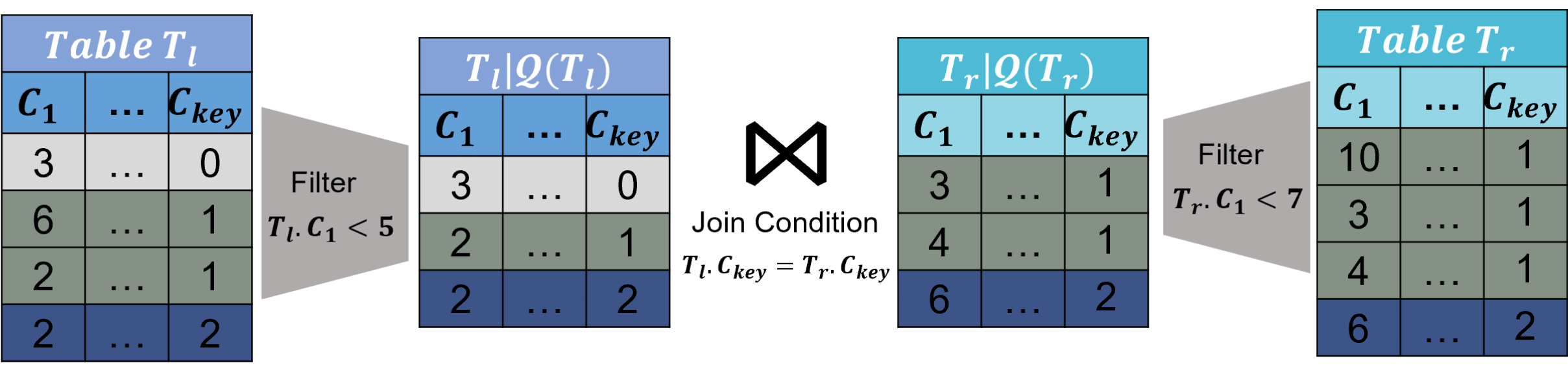}
    \caption{Sample of DistJoin's cardinality estimation.}
    \vspace{-5pt}
    \label{fig:sample}
\end{figure}

\begin{figure*}[!htbp]
    \centering
    \includegraphics[width=0.8\linewidth]{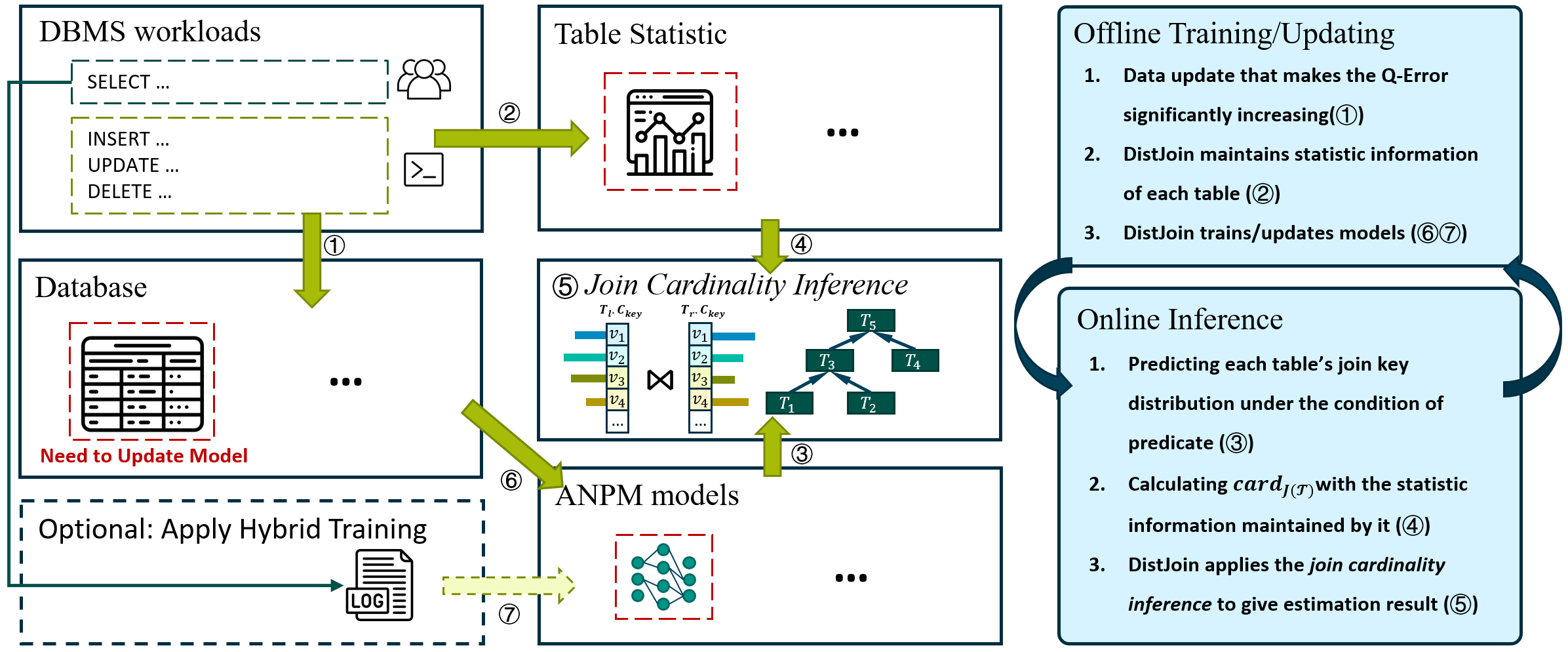}
    \caption{Overview of DistJoin.}
    \vspace{-15pt}
    \label{fig:overview}
\end{figure*}

The workflow of DistJoin is illustrated in \autoref{fig:overview}. Taking \autoref{fig:sample} as an example, DistJoin uses ANPM to estimate probability distributions (step \raisebox{-0.2ex}{\ding{194}}): $P_{T_l}(C_{key},C_1<5)=(\frac{1}{4},\frac{1}{4},\frac{1}{4}), P_{T_r}(C_{key},C_1<7)=(0,\frac{1}{2},\frac{1}{4}), P_{T_l}(C_{key})=(\frac{1}{4},\frac{1}{2},\frac{1}{4}), P_{T_r}(C_{key})=(0,\frac{3}{4},\frac{1}{4})$. Then we use the selectivity-based join cardinality inference to calculate the ratio of filtered tuples to the original join schema result by \autoref{equ:two_table_equi_join} :

\begin{equation}    
    P_{\mathcal{T}}(C_{key},Q)=\frac{\left[4 \cdot (\frac{1}{4},\frac{1}{4},\frac{1}{4})\right] \odot \left[4 \cdot (0,\frac{1}{2},\frac{1}{4})\right]}{\left[4 \cdot (\frac{1}{4},\frac{1}{2},\frac{1}{4})\right]\cdot \left[4 \cdot (0,\frac{3}{4},\frac{1}{4})\right]}=\left(0,\frac{2}{7},\frac{1}{7}\right)  
    \label{equ:equi_join_example}
\end{equation}

DistJoin then uses the maintained join key distribution and \autoref{equ:get_schema_card} to calculate $card_{J(\mathcal{T})}=1 \times 0+2 \times 3+1 \times 1=7$ (step \raisebox{-0.2ex}{\ding{195}}). Finally, the cardinality can be estimated by \autoref{equ:get_card}: $card = 7 \cdot \left(0+\frac{2}{7}+\frac{1}{7}\right)=3$.

For non-equi joins like ``$T_l$ Join $T_r$ on $T_l.C_{key} \leq T_r.C_{key}$'', DistJoin perform \textit{cumsum} on the distribution of left table in \autoref{equ:equi_join_example}, and obtain the join keys' selectivity $$P_{\mathcal{T}}(C_{key},Q)=\frac{\left[4 \cdot (\frac{1}{4},\frac{2}{4},\frac{3}{4})\right] \odot \left[4 \cdot (0,\frac{1}{2},\frac{1}{4})\right]}{\left[4 \cdot (\frac{1}{4},\frac{3}{4},1)\right]\cdot \left[4 \cdot (0,\frac{3}{4},\frac{1}{4})\right]}=\left(0,\frac{4}{13},\frac{3}{13}\right)$$ Similarly, the cardinality of the join schema is 13, so the estimated cardinality is 7.

DistJoin's design differs fundamentally from existing methods:

\textbf{Comparison with NeuroCard:} While NeuroCard couples all tables and fanout values to estimate full outer join view probabilities, DistJoin preserves inter-table correlations through key distributions by maintaining lightweight autoregressive models and statistical information per table. This enables independent, on-demand model updates when individual tables' data changes.

\textbf{Comparison with FactorJoin:} As shown in \autoref{equ:factor_equi_join}, FactorJoin employs binning to handle large join key domains, sacrificing accuracy. In contrast, DistJoin's efficient distribution estimation directly computes complete join key distributions, eliminating binning errors and supporting equi joins, non-equi joins, and outer joins with high accuracy. Additionally, DistJoin calculates join query selectivity using estimated distributions and multiplies by efficiently computed join schema cardinality $card_{J(\mathcal{T})}$, mitigating variance accumulation - a key improvement over FactorJoin's direct calculation approach.
\begin{align}
\label{equ:factor_equi_join}
%\scalebox{0.9}{\mbox{$\displaystyle 
\hat{\text{card}} = \sum_{\substack{v \in \text{bin}( \\ \mathcal{D}(T_i.C_{\text{key}}))}} 
\prod_{i=1}^N \hat{p}_{T_i}(T_i.C_{\text{key}}=\text{bin}(v), \mathcal{Q}(T_i)) |T_i|
%$}}
\end{align}

The remainder of this paper details DistJoin's join key distribution estimation in \autoref{section:dist_estimation} and its \textit{selectivity-based join cardinality inference} in \autoref{section:joincard_est}.

\section{Join Cardinality Inference Based On Single Tables' Distribution}
\label{section:joincard_est}
This section first analyzes the variance accumulation issue in FactorJoin's join cardinality inference method~\autoref{equ:factor_equi_join}, then introduces DistJoin's \textit{selectivity-based join cardinality inference} approach.

\subsection{Previous Method's Variance Accumulation Problem}
\label{subsection:FactorJoinVarAnalysis}
Although FactorJoin is capable of providing unbiased results given unbiased probability estimates, the formula in \autoref{equ:factor_equi_join} suffers from severe variance accumulation, where variance scales with the square of the Cartesian product cardinality of joined tables (denoted by $\text{card}_{J(\mathcal{C})} = \prod_{i=1}^n|T_i|$).

\begin{theorem} \label{thm:FactorJoinVariance}
Given the join cardinality inference method defined in \autoref{equ:factor_equi_join}, let $\text{card}$ be the true cardinality and $p_i(v,\mathcal{Q})$ be the true probability, the expectation and variance of the estimation satisfy the following formula under the assumption of unbiased probability estimation for individual tables:
\vspace{-2pt}
\begin{align}
  \mathbb{E}[\hat{\text{card}}] &= \sum_{v}\prod_{i=1}^{N} |T_i|p_i(v,\mathcal{Q}) = \text{card} \label{equ:FactorJoinExp} \\
  \text{Var}(\hat{\text{card}}) &\propto \prod_{i=1}^{N}|T_i|^2
\end{align}
\vspace{-10pt}
\end{theorem}

This \autoref{thm:FactorJoinVariance} is proved in the \textbf{Appendix}.
\vspace{-10pt}
\subsection{DistJoin's Equi Join Cardinality Inference}
As demonstrated, $\prod_{i=1}^n|T_i|^2$ and $\text{Var}(\hat{card})$ grow rapidly with increasing join tables and table cardinalities. This issue arises from the accumulation of errors in the estimated distribution during the recursive computation of join key counts. 

To address this, we propose a novel \textit{selectivity-based join cardinality inference} approach. Intuitively, we recursively compute the proportion of join key count in the join schema result instead of calculating the join key count itself. We then compute the real cardinality of the join schema based on the maintained join key distribution. Finally, we can estimate the queries' cardinality by multiplying the proportion by the join schema's true cardinality. On the one hand, the error is mitigated by the accurate cardinality of the join schema, as this correctness provides additional valuable information. On the other hand, the proportion's value is much smaller than the join key's counts, so the error accumulation problem is less serious. Besides, compared to weighted averaging or Bayesian combination, such an approach can obtain the true cardinality if the estimated distribution has no error, which makes it a better choice.

This method recursively computes, from the bottom up along the join plan tree, the proportion of each join key value under predicate filtering in the unfiltered join result $J(\mathcal{T})$. Summing these proportions yields the overall query selectivity. Combined with the maintained key column distribution information used to calculate the true cardinality $\text{card}_{J(\mathcal{T})}$ via \autoref{equ:get_schema_card}, multiplying selectivity by this cardinality produces the final estimate. The process is formalized as:
\begin{align}
    &\hat{p}_{T_l \Join T_r}(C_{key}=v, \mathcal{})\\
    &= \frac{\hat{p}_{T_l}(T_l.C_{key}=v,\mathcal{Q}(T_l))  \hat{p}_{T_r}(T_r.C_{key}=v,\mathcal{Q}(T_r))}{\sum\hat{p}_{T_l}(T_l.C_{key}=v) \hat{p}_{T_r}(T_r.C_{key}=v)} \label{equ:two_table_equi_join}\\
    &\hat{p}_{\mathcal{T}}(C_{key}=v,\mathcal{Q}) \\ 
    &= \frac{\hat{p}_{\mathcal{T}_{l}}(\mathcal{T}_{l}.C_{key} = v, \mathcal{Q}) \hat{p}_{\mathcal{T}_{r}}(\mathcal{T}_{l}.C_{key} = v, \mathcal{Q})}{\sum_{v \in \mathcal{D}(\mathcal{T})}{\hat{p}_{\mathcal{T}_{l}}(\mathcal{T}_{l}.C_{key} = v) \hat{p}_{\mathcal{T}_{r}}(\mathcal{T}_{l}.C_{key} = v)}} \label{equ:multi_tables_equi_join}
\end{align}
\begin{align}
    &\mathcal{T}_{l} = \{T_1, \dots, T_{m}\}, \mathcal{T}_{r} = \{T_{m + 1}, \dots, T_N\} \label{equ:split_rules}\\
    &\text{card}_{J(\mathcal{T})} = \sum\prod_{i=1}^Np_{T_i}(T_i.C_{key}=v)|T_i| \label{equ:get_schema_card}\\
    &\hat{\text{card}} = \text{card}_{J(\mathcal{T})} \sum_{v \in \mathcal{D}(\mathcal{T})}\hat{p}_{\mathcal{T}}(\mathcal{T}.C_{key}=v, \mathcal{Q}) \label{equ:get_card}
\end{align}
%_{v \in \mathcal{D}(T_l.C_{key})\cup \mathcal{D}(T_r.C_{key})}
% _{v \in \cup_{i=1}^N\mathcal{D}(T_i.C_{key})}
Here, $\mathcal{T}$ denotes multiple tables' join result, $card_{J(\mathcal{T})}$ represents the join schema cardinality on $\mathcal{T}$, and $p_{T_i}(T_i.C_{key})$ indicates the true distribution probability of the join key column. \autoref{equ:two_table_equi_join} presents the base case for two-table joins, while \autoref{equ:multi_tables_equi_join} and \autoref{equ:split_rules} define the recursive formulas. \autoref{equ:get_schema_card} calculates the join schema cardinality, which DistJoin caches for reuse. Finally, \autoref{equ:get_card} estimates cardinality by multiplying the summed selectivity by $\text{card}_{J(\mathcal{T})}$.

Notably, DistJoin avoids conditional independence assumptions despite \autoref{equ:two_table_equi_join} and \autoref{equ:multi_tables_equi_join}'s apparent similarity to such assumptions because \autoref{equ:two_table_equi_join} is the result of eliminating $|T_l||T_r|$ from both the numerator and denominator, which is a precise calculation of each key's selectivity based on key counts.

Under the assumptions in \autoref{assump:error_distribution} and ~\autoref{assump:DistJoinAnalysis}, this approach provides unbiased estimates with variance proportional to $\text{card}_{J(\mathcal{T})}^2$, as proven in \autoref{section:error_analysis}.

\textbf{\underline{Inference Efficiency.}} Leveraging ANPM's high estimation efficiency, this process avoids the computational expense suggested by the formulas. ANPM estimates complete join key column distributions, enabling full utilization of GPU parallel computation or CPU AVX instructions. For two-table joins, DistJoin employs vectorized calculation:
\begin{align}
\vspace{-15pt}
    \hat{\mathbf{P}}_{\mathcal{T}}(C_{key},\mathcal{Q}) &= \frac{\hat{\mathbf{P}}_{\mathcal{T}_{l}}(C_{key}, \mathcal{Q}) \odot \hat{\mathbf{P}}_{\mathcal{T}_{r}}(C_{key}, \mathcal{Q})}{\hat{\mathbf{P}}_{\mathcal{T}_{l}}(C_{key}) \cdot \hat{\mathbf{P}}_{\mathcal{T}_{r}}(C_{key})}
    \vspace{-15pt}
\end{align}

Here, denominator vectors follow consistent distinct value encoding and are recursively calculated using the same formula with $\mathcal{Q}=\emptyset$. The algorithm appears in \autoref{alg:InferenceForJoinTypes}. DistJoin also supports outer joins by assigning probability 1 to non-existent join key values in corresponding tables before performing equi-joins.

\vspace{-15pt}
\subsection{Expectation and Variance Analysis}
\label{section:error_analysis}

This subsection presents an expectation and variance analysis for equi joins, comparing DistJoin's performance with FactorJoin's approach.

\begin{theorem} \label{thm:DistJoinExpVar}
Let $card$ denote the true cardinality and $p_T(\cdot)$ represent the true probability. The expectation and variance of DistJoin's equi join cardinality inference process satisfy:
\begin{align}
    \mathbb{E}[\hat{\text{card}}] &\approx \text{card}_{J(\mathcal{T})} \frac{\sum_v p_{T_l}(v,\mathcal{Q})p_{T_r}(v,\mathcal{Q})}{\sum_v p_{T_l}(v)p_{T_r}(v)} = \text{card} \\
    \text{Var}(\hat{\text{card}}) &\propto \text{card}_{J(\mathcal{T})}^2
\end{align}
\end{theorem}

We proved the \autoref{thm:DistJoinExpVar} with the \autoref{assump:DistJoinAnalysis} in the \textbf{Appendix}.

A comparative analysis of variance between FactorJoin and DistJoin, formalized in \autoref{thm:FactorJoinVariance} and \autoref{thm:DistJoinExpVar}, reveals fundamental insights:

\begin{enumerate}[label=\alph*)]
    \item \textbf{Unbiased Estimation}:
    \begin{itemize}
        \item \textit{FactorJoin-style approaches} achieve unbiased estimation under standard independence assumptions (\autoref{assump:error_distribution})
        \item \textit{DistJoin} requires stronger regularity conditions specified in \autoref{assump:error_distribution} and \autoref{assump:DistJoinAnalysis}
    \end{itemize}
    
    \item \textbf{Variance Scaling}:
    \begin{itemize}
        \item DistJoin's variance scales with the square of the \textit{join schema cardinality} $\text{card}_{J(\mathcal{T})} = \sum_{v}\prod_{i}p_{T_i}(T_i.C_{key}=v)|T_i|$), with variance suppression through normalization factor $\hat{S}$.
        \item FactorJoin's variance scales with the square of the \textit{Cartesian product cardinality} ($\text{card}_{J(\mathcal{C})} = \prod_{i=1}^n |T_i|$)
    \end{itemize}
    
    \item \textbf{Cardinality Growth Dynamics}:
    Assuming the coefficient term dominated by variance in \autoref{equ:FactorJoinVariance} and \autoref{equ:DistJoinVariance} is significantly smaller than the baseline value, we have:
    \begin{equation}
        \frac{\text{card}_{J(\mathcal{C})}}{\text{card}_{J(\mathcal{T})}} = \Omega\left( \frac{\prod_{i=1}^n |T_i|}{|\Join_{i=1}^n T_i|} \right)
    \end{equation}
    This ratio grows significantly with an increasing number of join tables $(n)$.
\end{enumerate}

This analysis demonstrates DistJoin's inherent advantage in large-scale join scenarios through its tighter variance bound. Controlled experiments hybridizing DistJoin's architecture with FactorJoin's inference methodology empirically validate this theoretical advantage. Results consistently show that DistJoin's selectivity-driven approach achieves lower error than FactorJoin and hybrid baselines in multi-table joins.

\subsection{DistJoin's Non-Equi Join Cardinality Inference}

This subsection details DistJoin's support for non-equi joins. Consider $T_l \text{ join } T_j \text{ on } T_l.C_{key}\leq T_r.C_{key}$ as an example. For $T_r.C_{key}=v$, it joins with $T_l.C_{key} \leq v$. Thus, we calculate:
\begin{align}
    \hat{p}_{T_r}(T_l.C_{key}=v, \mathcal{Q}) \sum_{k=1}^{|\mathcal{D}(T_l.C_{key})|} \hat{p}_{T_l}(T_l.C_{key}=k, \mathcal{Q})
\end{align}

Traversing $v$ from 1 to $|\mathcal{D}(T_l.C_{key})|$ enables optimization of repeated sum calculations through the \textit{cumsum}() function applied to $\hat{\mathbf{P}}_{T_r}(T_r.C_{key}, \mathcal{Q})$. For join conditions like $<$, we right-shift the distribution vector, pad with zero at the beginning, and drop the last item before performing \textit{cumsum}(). The recursive inference process for all supported join conditions appears in \autoref{alg:InferenceForJoinTypes}.

\begin{figure}[!htbp]
    \centering
    \includegraphics[width=0.9\linewidth]{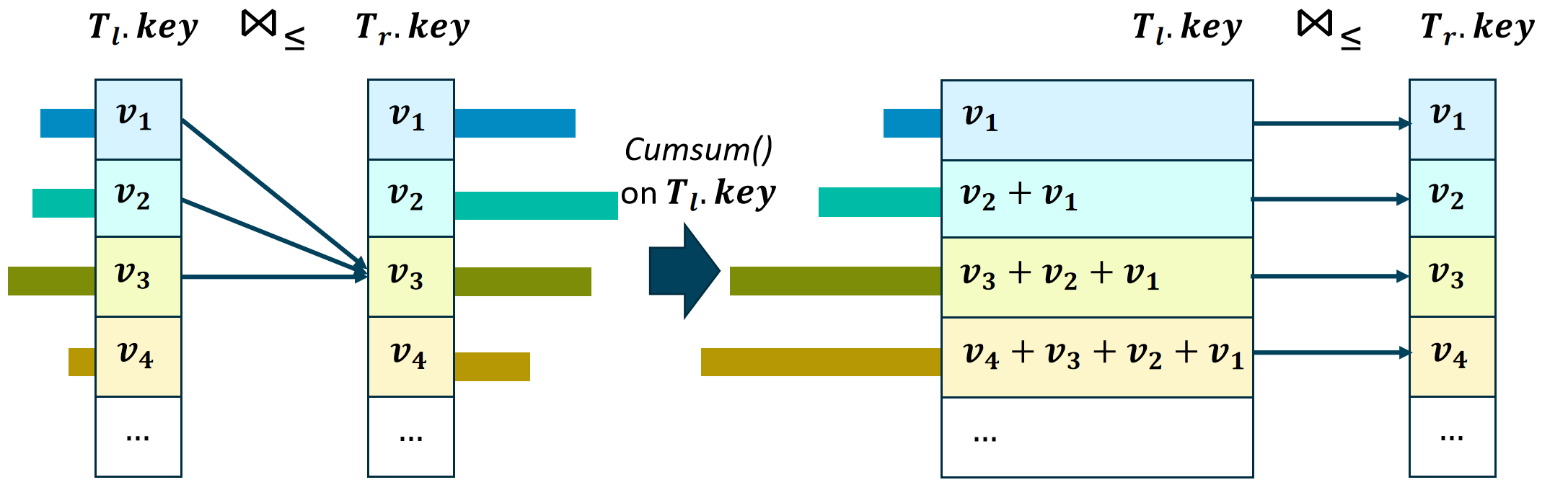}
    \caption{Illustration of DistJoin's deal with non-equi joins.}
    \vspace{-5pt}
    \label{fig:Non_equi_join}
\end{figure}

Non-equi joins typically achieve higher accuracy than equi joins due to each source key having more target keys to join, resulting in higher selectivity. The \textit{cumsum} operation virtually eliminates the possibility of serious underestimation caused by incorrectly estimating a key's probability as 0, except for the first join key value during \textit{cumsum} calculation.

\section{Individual Table's Distribution Estimation}
\label{section:dist_estimation}

This section introduces our Adaptive Neural Predicate Modulation and how DistJoin employs it to estimate each table's key distribution $\hat{P}_{T_i}(T_i.C_{key}, \mathcal{Q}(T_i))$.

\begin{figure}[t]
    \centering
    \includegraphics[width=0.9\linewidth]{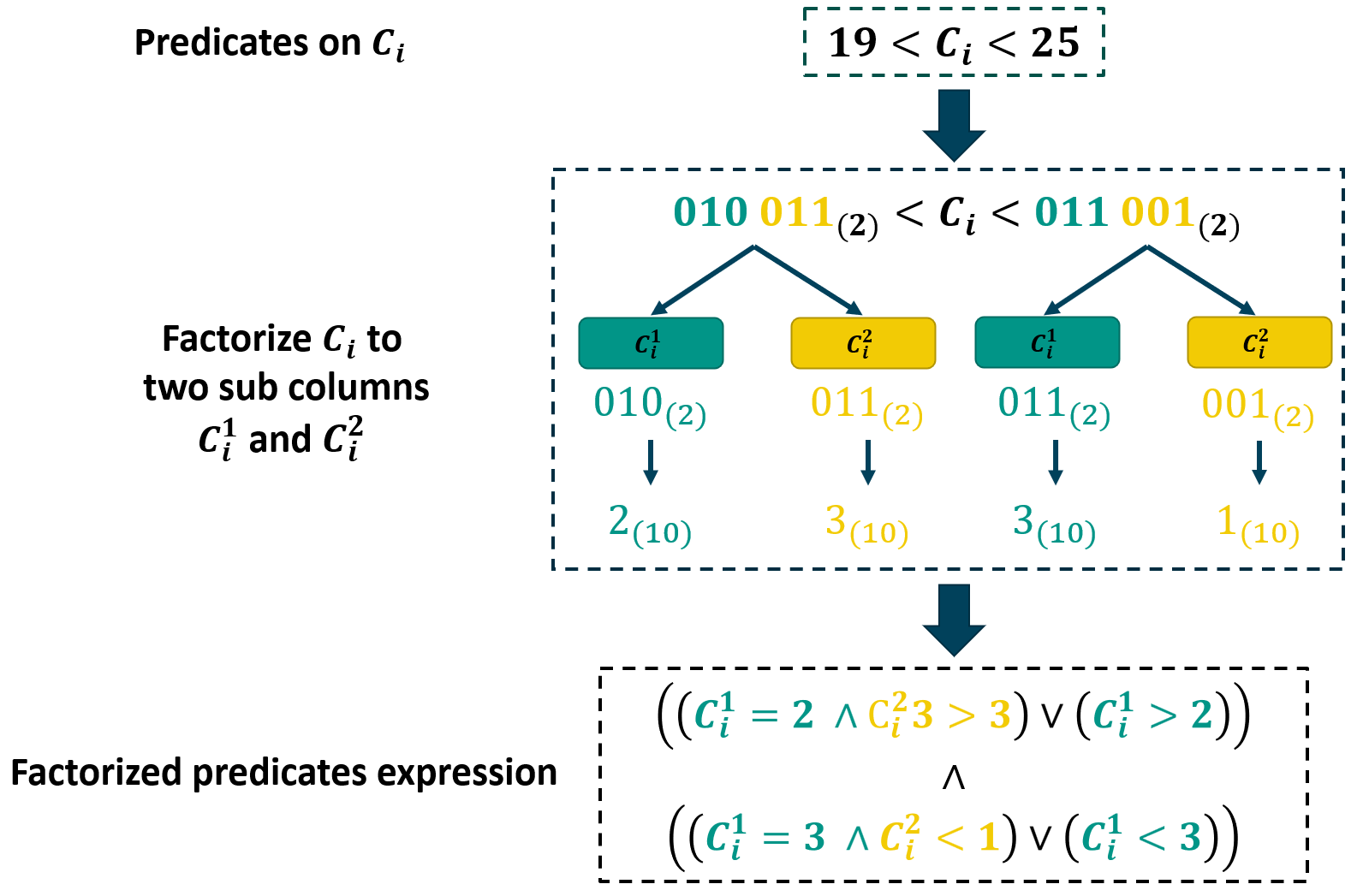}
    \caption{Example of factorized predicates.}
    \label{fig:pred_factorize_a}
\end{figure}

\subsection{Predicates Lossless Factorization}
\label{subsection:pred_factorization}
Naru~\cite{Naru} introduced a significant advancement by formulating single-table cardinality estimation as an autoregressive process. However, its reliance on progressive sampling for range queries necessitates thousands of samples and multiple neural network forward passes per estimation. This approach becomes infeasible for join key columns where $|\mathcal{D}(T_i.C_{key})|$ can reach $10^8$, as estimating $\hat{P}_{T_i}(T_i.C_{key}, \mathcal{Q}(T_i))$ would require storing and processing a multi-dimensional array exceeding $R^{2k\times 1e8}$ dimensions (with typical sampling size $k=2000$). BayesCard~\cite{BayesCard} similarly struggles with estimating probabilities for millions of distinct join key values within practical time constraints. These limitations force existing methods to adopt compromise techniques like binning and sampling, despite their drawbacks in accuracy, non-equi join support, and update flexibility.

DistJoin builds upon our previous single-table estimator Duet~\cite{Duet}, detailed in \autoref{subsection: Preliminary}. While Duet eliminates sampling and predicts complete probability distributions in a single forward inference, its inference time and memory costs remain high for large domain sizes due to high-dimensional embedding layers and output distributions - a common challenge for join key columns. 

NeuroCard's lossless column factorization technique~\cite{NeuroCard} addresses this by converting columns with large domains into binary representations and splitting them into sub-columns. This reduces domain size from $|\mathcal{D}(C)|$ to less than $m\times 2^{\lceil\frac{\lfloor log_2(|\mathcal{D}(C)|)\rfloor+1}{m}\rceil}$, where $m$ is the number of sub-columns. However, this technique is incompatible with Duet's predicate-based input approach.

As shown in \autoref{fig:pred_factorize_a}, applying column factorization to predicate values significantly increases the complexity of the predicate expression. For instance, the predicate $19 < C_i < 25$ has a simple boundary when unfactorized but becomes a complex nested logical expression after factorization, as illustrated in \autoref{fig:pred_factorize_a}. The filtered region remains a hypercube, and its boundary becomes a complex piecewise hyperplane, with complexity growing with the sub-column 
count.

%\autoref{fig:pred_factorize_b} demonstrates that while 

For a predicate $C_i > v$, let $k_i$ denote the number of sub-columns from $C_i$'s factorization, and let $C_i^j$ and $v_i^j$ represent the $j$-th factorized column and predicate value respectively (with $j=0$ as the highest bit). The factorized predicate expression becomes:
\begin{equation}
\label{equ:full_factorized_preds}
( C_i^0 > v_i^0 ) \ \lor \ \dots \ \lor \ ( C_i^0 = v_i^0 \land \dots \land C_i^{k_i}= v_i^{k_i} \land C_i^{k_i} > v_i^{k_i} )
\end{equation}

To address this challenge, we embed both original and factorized predicate information into the input vector. As shown in \autoref{equ:full_factorized_preds}, predicate factorization follows a consistent pattern: $n_i$ conjunctive clauses connected by disjunctions, where only the final term's predicate operator matches the original predicate while preceding terms use equality operators. This pattern enables lossless value factorization without directly encoding complex logical expressions. For example, the predicate $C_i > v_i$ factorizes into $(C_i^1 > v_i^1, \dots, C_i^{k_i} > v_i^{k_i})$, where $v_i^j$ represents the factorized value corresponding to each sub-column.

For multiple predicates on the same original column, as illustrated in \autoref{fig:model_architecture}, DistJoin first factorizes and embeds each column's predicates independently. Following Duet's approach, it then sums the predicate embedding vectors for each sub-column to form the input for our ANPM model, which we detail in the following subsection.

\subsection{Adaptive Neural Predicate Modulation}
\begin{figure*}[h]
    \centering
    \includegraphics[width=0.8\linewidth]{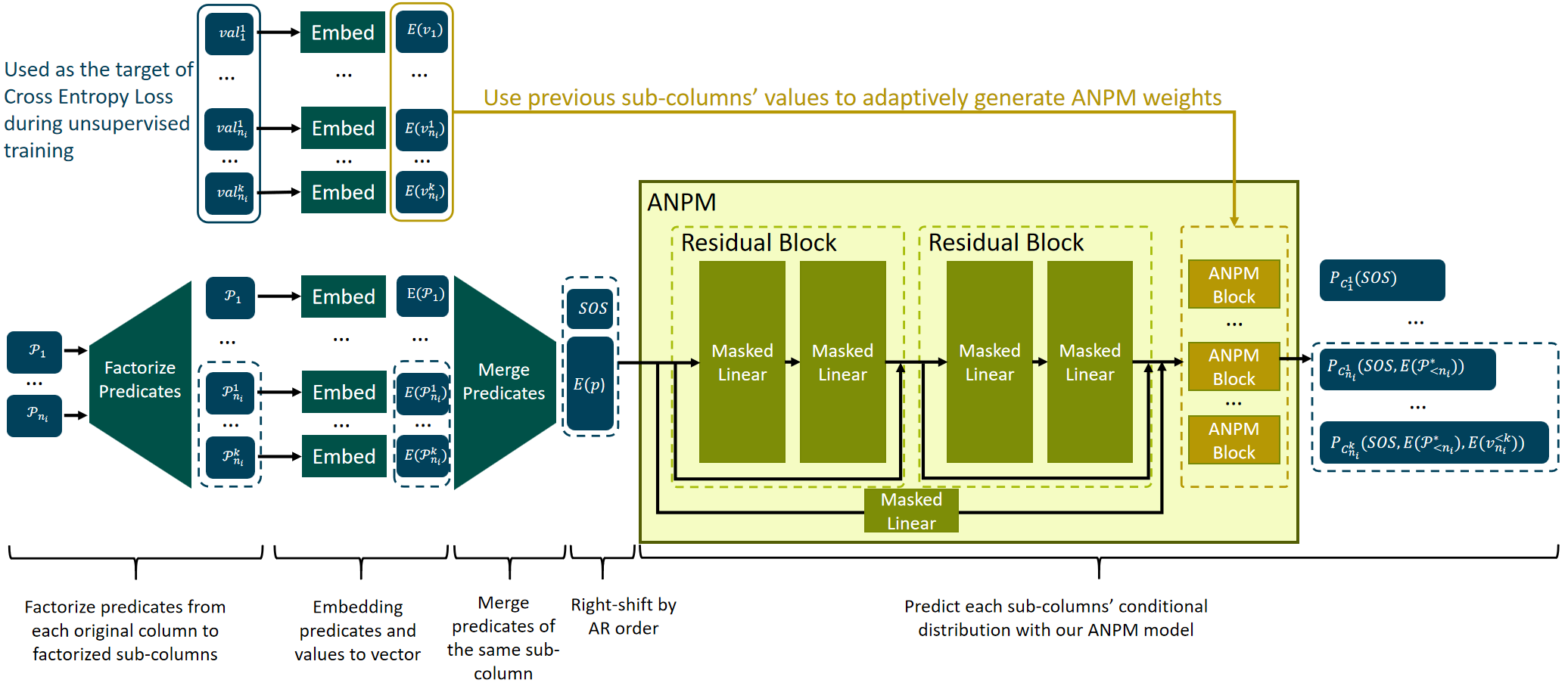}
    \caption{Architecture of Single Table's Distribution Estimator. }
    \vspace{-15pt}
    \label{fig:model_architecture}
\end{figure*}

The predicate factorization method provides lossless predicate semantics for $C_i$ (or its sub-columns) based on preceding original columns $C_{<i}$. However, for $C_i^j$, this approach alone cannot handle factorized predicate constraints from preceding $j-1$ sub-columns.

From \autoref{equ:full_factorized_preds}, we observe that the $j$-th sub-column $C_i^j$'s probability distribution depends on factorized predicates from the first $j-1$ sub-columns of the same original column. Taking $C_i > v_i$ as an example, simply making the $j$-th sub-column's distribution depend on these predicates would yield $\hat{P}(C_i^j|C_{i}^{1}>v_i^1 \land \dots, \land C_{i}^{j-1}>v_i^{j-1} )$, which is incorrect since conditions should follow \autoref{equ:full_factorized_preds}, leading to selectivity underestimation. \textbf{The core issue lies in common autoregressive models (e.g., Transformer, MADE) only representing conjunctive relationships through autoregressive masks, unable to handle complex logical expressions involving disjunctions within one inference.}

Another challenge is reconstructing original column distributions from sub-column distributions for selectivity estimation using \autoref{eq:selection} and \autoref{eq:indicator}. DistJoin reconstructs the original column's distribution through Algorithm~\ref{alg:OriginalColumnDistributionReconstruction}. \textit{Line 3} applies element-wise product between distribution $\mathbf{\hat{P}}(C_i^{<j}| \mathcal{P}_{<i})$ and estimated multi-dimensional distribution $\mathbf{\hat{P}}(C_i^j| C_i^{<j}, \mathcal{P}_{<i}) \in R^{\prod_1^{n}|\mathcal{D}(C_i^j)|}$ (where $\mathcal{P}$ is predicate). Intuitively, since the sub-columns are arranged from high to low bits~\cite{NeuroCard}, the column dimension of the element-wise product result represents the distinct value combinations of lower-bit sub-columns, while the row dimension represents all possible permutations of distinct values from higher-bit sub-columns. By unfolding the result in row-major order, we obtain the probability distribution of all distinct values for the original column. This lossless reconstruction requires estimating $\hat{P}(C_i^1, pred_{<i}) \in R^{|\mathcal{D}(C_i^1)|},\dots,\hat{P}(C_i^n|C_i^{<n}, pred_{<i})\in R^{\prod_1^n{|\mathcal{D}(C_i^j)|}}$, necessitating $n$ inference steps with up to $\prod_1^n{|\mathcal{D}(C_i^j)|}$ inputs per inference for each original column. This approach is infeasible due to excessive time and GPU memory requirements during training.
\begin{algorithm}[htbp]
    \caption{Lossless Reconstruction for Original Column's Distribution}
    \label{alg:OriginalColumnDistributionReconstruction}
    \LinesNumbered
    \footnotesize
    \KwIn{Sub-columns' predicate distributions:$\mathbf{\hat{P}}$}
    \KwOut{Reconstructed original column's distribution}
    $\mathbf{Dist} \longleftarrow \mathbf{\hat{P}}(C_i^1|\mathcal{P}_{<i})$; \\
    \For{$j \in [2, n]$}{
        $\mathbf{Dist} \longleftarrow$ flatten($\mathbf{Dist} \text{.unsqueeze(1)} \odot \mathbf{\hat{P}}(C_i^{j}|C_i^{<j}, \mathcal{P}_{<i})$); \qquad // $\mathbf{Dist} = \mathbf{\hat{P}}(C_i^{\leq j}| \mathcal{P}_{<i})$ here \\
    }
    $\mathbf{Dist\_Slice} \longleftarrow \mathbf{Dist}[:|\mathcal{D}(C_i)|]$; \qquad // Drop non-existent combinations \\
    $\mathbf{Dist} \longleftarrow \mathbf{Dist}\text{.sum()}/\mathbf{Dist\_Slice}\text{.sum()}$; \qquad // Normalization \\
    \Return $\mathbf{Dist}$; \qquad //$\mathbf{\hat{P}}(C_i|\mathcal{P}_{<i})$ 
\end{algorithm}

\begin{figure*}[h]
    \centering
    \includegraphics[width=0.8\linewidth]{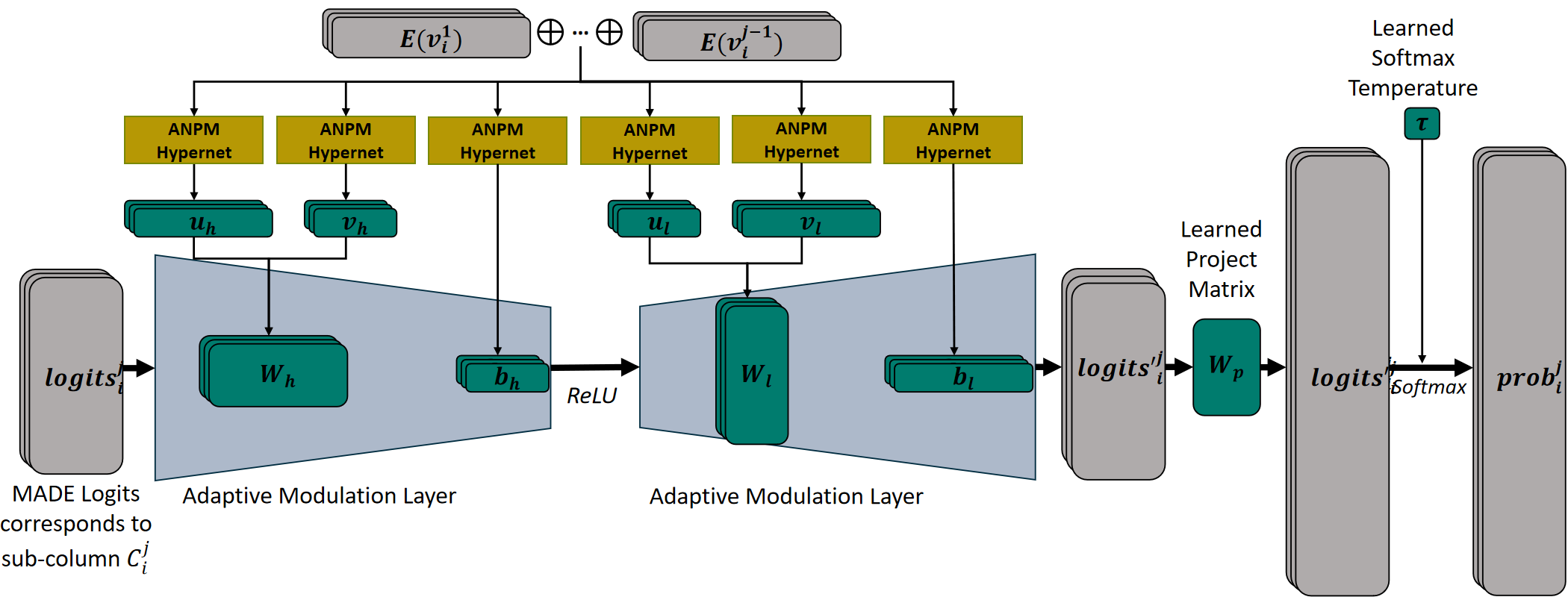}
    \caption{Architecture of the ANPM Block.}
    \label{fig:ANPM_architecture}
    \vspace{-15pt}
\end{figure*}

To address these challenges, we designed the ANPM model, illustrated in \autoref{fig:model_architecture} and \autoref{fig:ANPM_architecture}. The ANPM comprises a ResMADE network~\cite{MADE,ResMADE,Naru}, six ANPM HyperNetworks per factorized column, and a learned \textit{softmax} temperature coefficient.

The ResMADE network's autoregressive order and masks are carefully designed. From the perspective of the original columns, the order is arbitrary, except that the join key column must be last. From the perspective of factorized columns, sub-columns from the same original column are placed adjacent, ordered from higher to lower bits. We address a fitting capability limitation in previous works~\cite{Naru, NeuroCard} where the first distribution predicted by MADE (or ResMADE)~\cite{MADE, ResMADE} was independent of inputs, relying solely on the last linear layer's bias. Inspired by NLP techniques, we introduce an \textit{SOS} (start of sequence) token at the beginning of the input sequence and right-shift the input sequence order to improve the accuracy of the first column distribution prediction. The \textit{SOS} token remains at the autoregressive sequence's start.

Autoregressive masks adhere to constraints where output corresponding to $C_i^j$ depends on any $E(\mathcal{P}_{<i}^*)$ predicate inputs but not $E(\mathcal{P}_{\geq i}^*)$, maintaining autoregressive dependency constraints from the original columns' perspective. Thus, ResMADE's output $logits_i^j$ is predicted based on preceding $i-1$ original columns' predicates.

To balance complex non-linear inter-column correlations with model training and inference speed, we implement a learned mixed activation layer. This layer applies different activation functions to ResMADE's final layer output and outputs their weighted sum, enhancing the model's representational capacity. To further adapt the high skewness column, we employ a learned temperature coefficient for each sub-column's softmax activation function.

For factorized predicates, we modify Duet's approach: instead of predicate-based input for same-original-column sub-columns, we embed true values of the first $j-1$ sub-columns and input them into the ANPM Block during training, similar to Naru. As shown in \autoref{fig:ANPM_architecture}, the ANPM Block employs six lightweight HyperNetworks (2-layer MLPs with 64-unit width) to predict low-rank decomposition results for weight matrices and bias vectors. Matrix multiplication reconstructs weights for two linear adaptive modulation layers, embedding the first $j-1$ sub-columns' true values into their weights and biases. By sizing these layers appropriately, they form a bottleneck MLP, further optimizing calculations. This MLP modulates information from the first $j-1$ sub-columns into $logits_i^j$ during transformation.

ANPM's calculation for $C_i^j$ follows these formulas, with distinct embedding approaches for training and inference. As shown in \autoref{equ:concat_training}, ANPM concatenates factorized true value embedding vectors during training. During inference, ANPM generates combinations of the first $j-1$ sub-columns' distinct values along the batch dimension, avoiding dimensional explosion as shown in \autoref{equ:concat_inference}.

Finally, ANPM calculates the distribution as $\hat{P}(C_i^j|C_i^1=v_i^1,\dots,C_i^{j-1}=v_i^{j-1}, \mathcal{P}_{<i})\in R^{(\text{batch\_size},|\mathcal{D}(C_i^j)|)}$ during training and $\hat{P}(C_i^j|C_i^1,\dots,C_i^{j-1}, \mathcal{P}_{<i}) \in R^{\prod_1^j |\mathcal{D}(C_i^j)|}$ during inference.

The key advantage of ANPM lies in its ability to decouple predicate processing from factorized value combination handling. By employing dedicated ANPM blocks for each sub-column, the ResMADE component focuses exclusively on processing predicate information. The ANPM blocks efficiently handle all possible combinations of factorized values from $C_i^{<j-1}$. This architecture enables ResMADE, the most computationally intensive component, to perform parallel inference across all sub-columns during the column distribution reconstruction phase, where the model must predict $\hat{P}(C_i^j|C_i^1,\dots,C_i^{j-1}, \mathcal{P}_{<i})$.

Without ANPM, ResMADE would be unable to simultaneously represent all sub-column values of $C_i$ and the factorized predicates in a single input. This limitation would necessitate processing each original column's sub-columns sequentially: feeding ResMADE with all factorized predicates from $C_{<i}$ while iterating through all possible values of $C_i^{<j-1}$. This approach, similar to NeuroCard's method, would require multiple large-scale inference passes, significantly compromising estimation efficiency.

\textbf{\underline{Scalability Analysis.}} Due to the generalization of Duet's training framework\cite{Duet}, the table's cardinality does not significantly limit DistJoin's performance. The real bottleneck is the domain size. Considering a table with a maximum domain size of $2^{30}$, we can divide it into three sub-columns with domain sizes of $2^{8}, 2^{8}, 2^{14}$. Since the training process is not dependent on the reconstruction process, the training process can remain efficient. The inference performance is limited by predicting the third sub-column's distribution under the inputs of dimension $2^8 \times 2^8$, which is $4\times$ larger than the biggest inputs in our experiments. Considering the maximum GPU memory utilization of our experiment is 3.66GB, a GPU with 16GB of memory can handle the domain size of $2^{30}$. By batching the inputs at the cost of inference speed, the maximum domain size it can handle is about $2^{32}$, which approaches the uint32 type limit commonly used for primary key IDs in some DBMSs. Thus, our method accommodates most database scenarios, such as online e-commerce~\cite{url1,url2} and social media~\cite{url3}. DistJoin is limited by domain size only in some scenarios, such as IoT and time series databases.

\textbf{\underline{Training.}} DistJoin employs Duet's method for training sample generation. Specifically, it randomly samples original and factorized tuples from the table, using Duet's predicate generation algorithm to create predicate inputs constraining these tuples. Factorized tuples serve dual purposes: as learning targets for cross-entropy loss (equivalent to NLL loss) and as inputs to the ANPM Block. Since the ANPM model trains through unsupervised learning, it's not a label leakage. This training approach on factorized tables avoids calculations involving the original column's large domain size.

\textbf{\underline{Inference.}} During inference, ANPM Blocks utilize embedding weights containing all distinct values' embedding vectors for each factorized column as inputs. The distinct values' dimension serves as the batch dimension. Although this approach may appear resource-intensive, ANPM Blocks' placement after ResMADE ensures that ResMADE processes only a single sample during inference. Furthermore, ANPM Blocks employ lightweight MLP networks with low-rank decomposition, maintaining high efficiency even with thousands of sub-column distinct values. This method also alleviates combinatorial explosion across different original columns. Experimental results confirm ANPM's efficiency.

Finally, we reconstruct each original column's distribution under preceding predicates using \autoref{alg:OriginalColumnDistributionReconstruction}. DistJoin then estimates individual table selectivity $\hat{p}(\mathcal{Q}(T))$ using \autoref{eq:selection} and \autoref{eq:indicator}, calculating $\hat{P}(C_{key}, Q(T))=\hat{P}(C_{key} | Q(T))\hat{p}(\mathcal{Q}(T))$.

%\FloatBarrier 

\section{Experiments} \label{section:Experiment}
This section presents a comprehensive evaluation of DistJoin against common baselines.

\subsection{Experimental Setting}
\textbf{\underline{Datasets \& Workloads.}} We evaluate DistJoin's performance using the IMDB dataset with JOB-light and JOB-light-ranges workloads, the most widely adopted benchmarks in cardinality estimation research. Given the lack of established workloads for learning-based cardinality estimators on non-equi joins, we extended JOB-light and JOB-light-ranges by replacing join conditions with non-equi joins and recalculating true cardinalities. \autoref{tab:IMDB} and \autoref{tab:Workloads} present statistical information that demonstrates the wide scenario coverage of these benchmarks. Such a large cardinality may appear in the intermediate results of complex aggregation queries in OLAP scenarios. As shown in \autoref{tab:IMDB}, the dataset exhibits diverse column numbers, types, and NDV ranges. Additionally, we constructed a pre-updated dataset by randomly sampling $80\%$ of tuples from each table, simulating pre-update conditions for data and model update evaluations. 

\autoref{tab:Workloads} and \autoref{fig:workload_dist} reveal that the $10$ test workloads cover extensive ranges of query true cardinality and join schema cardinality. Notably, JOB-light-ranges' non-equi joins exhibit approximately 4 orders of magnitude lower selectivity, presenting greater prediction challenges. Non-equi joins demonstrate significantly higher true cardinality, with the six tables' Cartesian product cardinality reaching $1.5e46$ - 13 orders of magnitude above the maximum $\text{card}_{J(\mathcal{T})}$. Consistent with \autoref{section:error_analysis}'s findings, we anticipate DistJoin's superior performance over FactorJoin. The JOB-light workload with non-equi operators shows substantially higher true cardinality ranges due to simpler predicates, effectively demonstrating query-driven baselines like MSCN's numerical challenges with non-equi joins. For query-driven baselines, we used ALECE's workload generation method and generated 100k queries that contain both equi and non-equi join queries.

\begin{table}[htbp]
\centering
\caption{Statistical information of tables in the IMDB dataset included in our workloads. 
$\mathcal{C}$ represents for categorical type and $\mathcal{N}$ represents for numerical type.}
\label{tab:IMDB}
\begin{tabular}{lllll}
\toprule
                & \#Rows      & \#Col & Col Types           & NDV Range          \\ \midrule
\textit{cast info}       & 360M        & 7         & $\mathcal{C}, \mathcal{N}$ & 11--23M            \\
\textit{movie companies} & 26M         & 5         & $\mathcal{C}, \mathcal{N}$ & 660K--2.6M         \\
\textit{movie info}      & 140M        & 5         & $\mathcal{C}, \mathcal{N}$ & 71--140M           \\
\textit{movie info idx}  & 13M         & 5         & $\mathcal{C}, \mathcal{N}$ & 5--13M             \\ 
\textit{movie keyword}   & 45M         & 3         & $\mathcal{C}$          & 470K--45M          \\
\textit{title}           & 25M         & 12        & $\mathcal{C}, \mathcal{N}$ & 1--25M             \\
\bottomrule
\end{tabular}
\vspace{-5pt}
\end{table}

\begin{table}[htbp]
\centering
\caption{Workloads statistical information.}
\label{tab:Workloads}
\resizebox{\linewidth}{!}{
\begin{tabular}{llllll}
\toprule
                                  & \#Col                      & \#Pred                        & Join & Card Range           & $\text{Card}_{J(\mathcal{T})}$ Range \\ \midrule
\multirow{5}{*}{JOB-light}        & \multirow{5}{*}{8}         & \multirow{5}{*}{1--5}         & =         & $9$--$9.5e9$                  & $4.1e6$--$6.7e11$                               \\
                                  &                            &                               & $>$       & $7.6e10$--$1.2e30$            & $4.7e13$--$1.1e32$                              \\
                                  &                            &                               & $<$       & $1.5e10$--$6.1e30$            & $4.4e13$--$1.9e32$                              \\
                                  &                            &                               & $\geq$    & $7.6e10$--$1.2e30$            & $4.7e13$--$1.1e32$                              \\
                                  &                            &                               & $\leq$    & $1.5e10$--$6.1e30$            & $4.4e13$--$1.9e32$                              \\ \midrule
\multirow{5}{*}{JOB-light-ranges} & \multirow{5}{*}{13}        & \multirow{5}{*}{3--6}         & =         & $1$--$2.3e11$                 & $4.1e6$--$6.7e11$                               \\
                                  &                            &                               & $>$       & $1.6e4$--$3.4e31$             & $4.7e13$--$1.1e32$                              \\
                                  &                            &                               & $<$       & $5.0e4$--$4.0e31$             & $4.4e13$--$1.9e32$                              \\
                                  &                            &                               & $\geq$    & $1.6e4$--$3.4e31$             & $4.7e13$--$1.1e32$                              \\
                                  &                            &                               & $\leq$    & $5.0e4$--$4.0e31$             & $4.4e13$--$1.9e32$                              \\ \bottomrule
\end{tabular}
}
\end{table}

\textbf{\underline{Baselines.}} We evaluate against the following baselines, covering conventional, data-driven, and query-driven methods.

\begin{enumerate}[label=\alph*)]
    \item \textbf{PG:} PostgreSQL (PG)~\cite{PG_URL}, an open-source database with one of the most accurate traditional cardinality estimators, serves as a common baseline. We obtain estimated cardinality using the SQL template: ``EXPLAIN (ANALYZE OFF) SELECT * FROM \textit{tables} WHERE \textit{join\_conditions} AND \textit{predicates};".
    
    \item \textbf{NeuroCard:} NeuroCard~\cite{NeuroCard}, a data-driven method based on a single autoregressive model trained on full outer join view samples. We use the paper's open-source code for evaluation.
    
    \item \textbf{MSCN:} MSCN~\cite{MSCN}, a classic query-driven method and common baseline, we use the version with sampling. The evaluation uses the paper's open-source code.

    \item \textbf{ALECE:} ALECE~\cite{ALECE}, an advance query-driven estimator proposed in recent years. We used the source code posted in the paper~\cite{ALECE} and modified its feature encoding to support non-equi join queries.

    \item \textbf{FactorJoin:} FactorJoin~\cite{FactorJoin}, an advance data-driven estimator proposed in recent years. The evaluation uses the paper's open-source code. 

    \item \textbf{UAE:} UAE~\cite{UAE}, a representative hybird method based on Naru and NeuroCard. 
\end{enumerate}

We employ recommended settings from each baseline's `readme' files, optimized by their respective authors.

\textbf{\underline{Metric.}} Following most cardinality estimation methods~\cite{DeepDB, Naru, NeuroCard, MSCN, FACE, FLAT,ASM}, we use Q-Error~\cite{QError} to evaluate the accuracy, an effective measure for CE problems~\cite{AreWeReady}:

\textbf{\underline{Hardware.}} All experiments run on a PC with:
\begin{itemize}
    \item CPU: AMD Ryzen9 7950X3D
    \item RAM: 32GB $\times 2$
    \item GPU: NVIDIA RTX4070Ti (12GB VRAM)
\end{itemize}

\subsection{Accuracy Evaluation}
\label{subsection:accuracy_evaluation}
We first compare DistJoin's accuracy with baselines. Results appear in \autoref{tab:comparison0}. Note that while estimated cardinality differs between ($>$,$\geq$) and ($<$,$\leq$), their Q-Error values diverge only beyond the fourth decimal place (except for MSCN). To avoid redundancy, we exclude $\geq$ and $\leq$ results from tables. We provide a comprehensive, intuitive comparison across all five join types in the \textbf{Appendix}. 

\begin{figure}[!htbp]
    \centering
    \includegraphics[width=\linewidth]{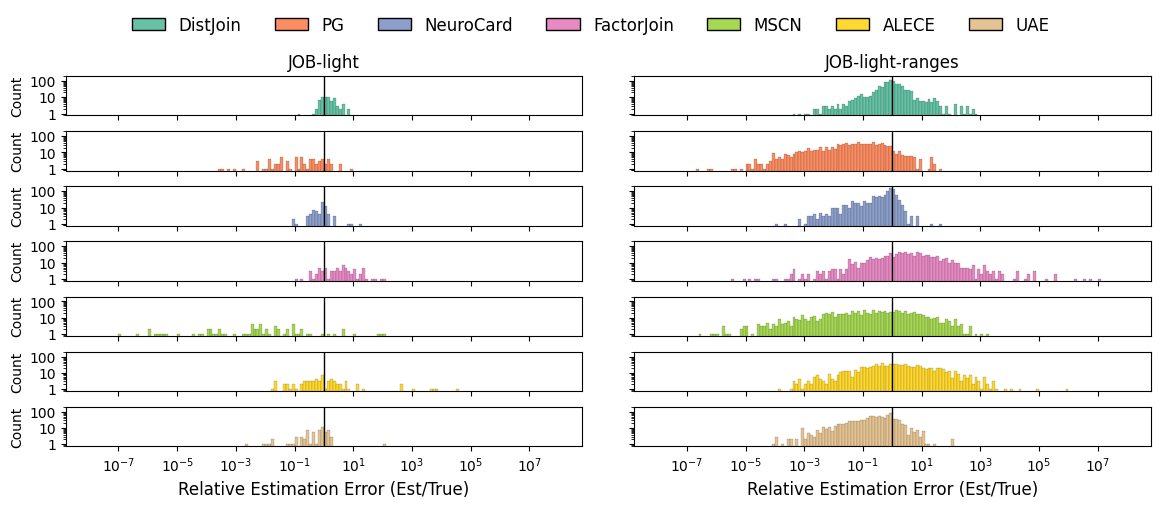}
    \caption{Distribution of the relative estimation error}
    \vspace{-5pt}
    \label{fig:r_error_dist}
\end{figure}

From these results, we draw the following conclusions:

\begin{enumerate}[label=\alph*)]
    \item \textbf{DistJoin outperforms baselines on almost all Q-Error percentiles.} NeuroCard's 50th Q-error is marginally lower than DistJoin's in both workloads under equi-joins, with this being the sole exception. However, DistJoin demonstrates significantly lower Q-Errors in other scenarios. Notably, DistJoin exhibits a lower 95th, 99th percentile, and maximum Q-Errors across all scenarios, indicating a less pronounced long-tail error distribution. Comparing DistJoin with FactorJoin validates our analysis in \autoref{thm:FactorJoinVariance} and \autoref{thm:DistJoinExpVar}.

    \item \textbf{DistJoin and ALECE exhibits low bias.} As shown in \autoref{fig:r_error_dist}, DistJoin and ALECE show significantly lower bias compared to other methods. Although FactorJoin theoretically has a low bias, its binning technique amplifies bias by violating \autoref{assump:error_distribution}.
    
    \item \textbf{MSCN and ALECE struggle with extreme cardinality ranges.} Both methods' non-equi join Q-Errors significantly exceed other methods. Compared to their accuracy on non-equi queries reported in their paper, the accuracy on equi-join queries is also significantly affected and drops when the training workload contains non-equi join queries with a wide cardinality range.
    
    \item \textbf{NeuroCard achieves slightly better median Q-Error but worse long-tail error distribution than DistJoin.} We attribute this to NeuroCard's training on full outer join view samples, which can span an enormous space~\cite{STATS_CEB}, and those invalid tuples sampled during its progressive sampling process.

    \item \textbf{FactorJoin's accuracy is constrained by its design.} As the first method of estimating join key distributions for join queries, FactorJoin made significant estimation compromises. Limited by BayesCard's~\cite{BayesCard} estimation throughput, FactorJoin's binning of join keys introduced additional errors, experimentally validating \autoref{subsection:FactorJoinVarAnalysis}'s variance analysis conclusions.

    \item \textbf{UAE's accuracy is limited under insufficient GPU resourcees.} While UAE outperforms NeuroCard in accuracy given sufficient GPU resources, its substantial training overhead~\cite{Duet} and our platform's limited GPU memory necessitated hyperparameter adjustments for timely training completion. Such resource constraints reflect real-world deployment challenges of the UAE, resulting in significantly reduced accuracy.
\end{enumerate}

In summary, the accuracy evaluation results demonstrate the limitations of the previous methods and DistJoin's superiority in both generality and accuracy. \textbf{DistJoin emerges as the only learned cardinality estimator effectively supporting non-equi joins among all baselines while achieving the highest accuracy in most scenarios}. This feature gives it a significant advantage in processing OLAP scenarios such as large-scale complex queries. 

\begin{table*}[htbp]
\centering
\caption{Q-Error comparison.}
\label{tab:comparison0}
\resizebox{0.9\textwidth}{!}{
\begin{tabular}{cccccccccccc}
\toprule
\multirow{2}{*}{Methods}    & \multirow{2}{*}{Join}      & \multicolumn{5}{c}{JOB-light}                                                                & \multicolumn{5}{c}{JOB-light-ranges}                                                     \\
\cmidrule(lr){3-7} \cmidrule(lr){8-12}
                            &                            & mean             & 50th          & 95th             & 99th             & max                 & mean            & 50th           & 95th            & 99th             & max               \\ \midrule
PG                          & \multirow{7}{*}{$=$}       & 165.00           & 6.77          & 838.32           & 3053.08          & 3630.94             & 9796.76         & 19.37          & 4547.06         & 61993.03         & $4.50e6$          \\
MSCN                        &                            & $2.11e5$         & 166.93        & $7.91e5$         & $4.29e6$         & $8.23e6$            & 11104.64        & 13.91          & 6358.10         & $1.40e5$         & $3.60e6$          \\
ALECE                       &                            & 847.09           & 3.41          & 2843.87          & 17172.43         & 39699.86            & 1015.15         & 7.00           & 488.92          & 2567.47          & $7.96e5$           \\ 
NeuroCard                   &                            & 2.42             & \textbf{1.30} & 9.15             & 14.27            & 18.47               & 33.10           & \textbf{1.75}  & 78.07           & 548.07           & 8169.00           \\
FactorJoin                  &                            & 10.03            & 4.09          & 36.06            & 88.94            & 102.60              & 23613.37        & 6.91           & 1447.98         & 70772.61         & $1.14e7$          \\
UAE                         &                            & 15.31            & 1.84          & 85.28            & 208.08           & 425.58              & 97.92           & 4.36           & 238.92          & 1734.42          & 10981.00          \\   
DistJoin                    &                            & \textbf{1.99}    & 1.49          & \textbf{4.53}    & \textbf{6.92}    & \textbf{7.63}       & \textbf{18.69}  & 1.94           & \textbf{43.21}  & \textbf{388.79}  & \textbf{2225.60}   \\
\midrule           
PG                          & \multirow{4}{*}{$>$}       & 2.74             & 1.62          & 6.59             & 12.53            & 22.38               & 9415.93         & 4.05           & 155.39          & 3209.22          & $7.10e6$           \\
MSCN                        &                            & $8.50e13$        & 97348.24      & $1.09e14$        & $2.18e15$        & $3.79e15$           & $6.38e16$       & 331.61         & $6.28e11$       & $2.06e15$        & $4.58e20$          \\
ALECE                       &                            & 1285.32          & 4.04          & 4372.08          & 30563.41         & 36295.48            & 52492.20        & 9.02           & 10878.93        & $1.29e5$         & $2.35e7$           \\ 
DistJoin                    &                            & \textbf{1.58}    & \textbf{1.12} & \textbf{4.30}    & \textbf{10.67}   & \textbf{10.94}      & \textbf{75.18}  & \textbf{1.57}  & \textbf{11.72}  & \textbf{79.92}   & \textbf{43969.74}  \\
\midrule           
PG                          & \multirow{4}{*}{$<$}       & 5.53             & 2.28          & 20.03            & 60.63            & 116.16              & 101.78          & 4.49           & 70.04           & 417.06           & 32506.44           \\
MSCN                        &                            & $3.28e14$        & 83835.69      & $3.33e14$        & $8.31e15$        & $1.49e16$           & $5.91e17$       & 595.61         & $1.63e12$       & $2.54e16$        & $4.58e20$          \\
ALECE                       &                            & 1204.29          & 13.19         & 6108.58          & 24046.73         & 41819.33            & 6940.42         & 12.01          & 6392.87         & $1.52e5$         & $1.46e6$           \\ 
DistJoin                    &                            & \textbf{1.30}    & \textbf{1.06} & \textbf{2.23}    & \textbf{3.98}    & \textbf{4.40}       & \textbf{4.16}   & \textbf{1.31}  & \textbf{7.68}   & \textbf{89.28}   & \textbf{338.58}    \\
\bottomrule
\end{tabular}
}
\vspace{-20pt}
\end{table*}

\subsection{Performance of Update and Inference}
Having demonstrated DistJoin's generality and accuracy regarding the ``Trilemma of Cardinality Estimation" in the last subsection, we now evaluate its update and inference speed against baselines on JOB-light-ranges with `$=$' join condition. We show maximum metric values across all tables for comparison, since they will be the performance bottleneck during training. For methods requiring complex pre-processing, we represent time as \textit{prepare time} + \textit{training time}.

For update speed, MSCN requires collecting training workload true cardinalities, NeuroCard needs full-outer join view samples, and FactorJoin must first apply its binning strategy to updated data. While MSCN and NeuroCard require complete retraining after data updates, FactorJoin theoretically only needs bucket updates rather than complete rebuilding. However, since this approach remains unimplemented in its code, we present its cold-start time alongside other methods. This limitation doesn't affect FactorJoin's status as the fastest updating method. Notably, DistJoin requires only distinct value sorting and encoding for data pre-processing, with training samples generated dynamically during training.

As shown in \autoref{tab:performanc_comparison}, FactorJoin achieves the fastest training and updating times, though its cold-start time is relatively longer due to binning. Proper optimization (e.g., replacing Python with C++) may significantly reduce pre-processing time. DistJoin ranks second in update speed, with its longest training time (310 seconds) occurring on the high-cardinality \textit{cast\_info} table. NeuroCard trains quickly but requires 150 seconds for full-outer join view sampling, limiting scalability. Additionally, its database-coupled model prevents flexible updates when only some tables change. The data annotation time of MSCN and ALECE exceeds 24 hours, making them nearly impractical for scenarios with frequent data updates or workload shifts.

Regarding inference time, MSCN's simple design makes it the fastest, though its high Q-Error diminishes this advantage. DistJoin's inference time is approximately $3\times$ FactorJoin's, a worthwhile tradeoff given its significant accuracy improvements. NeuroCard's inference time exceeds DistJoin's by over $2\times$ while delivering higher Q-Errors.

ALECE has the slowest inference speed among all tested methods. We attribute this to the significant memory overhead incurred by the default hyperparameters of the transformer model used when processing complex queries and large amounts of data. In some queries, the GPU memory exceeded the 12GB available on our experimental platform, resulting in data being swapped between GPU memory and main memory, which in turn slowed down the estimation speed.

In terms of model size, ALECE and DistJoin's models are significantly larger than others. As an autoregressive-based method, DistJoin's ANPM models are much larger than FactorJoin's BayesCard. Maintaining separate models for each table also significantly increases DistJoin's storage requirements compared to NeuroCard's. However, this storage overhead remains well below the maximum tolerable limits of database systems, imposing no storage pressure and compromising practicality.

Regarding GPU memory overhead, DistJoin requires more GPU memory than baselines during training. This stems from DistJoin's individual table estimators employing the same training framework as our previous work, Duet~\cite{Duet}. Training on more predicate samples helps reduce long-tail error distribution effects, prompting us to use a batch size of $16384$ to maximize training throughput, resulting in higher GPU memory overhead. However, during inference, DistJoin's GPU memory overhead is lower than NeuroCard's.

\begin{table}[htbp]
\centering
\caption{Performance comparison.}
% Since FactorJoin and DistJoin decouple models from the full outer join and maintain independent models per table, their models train separately. 
\label{tab:performanc_comparison}
\resizebox{\linewidth}{!}{

\begin{tabular}{lllllll}

\toprule
                          & MSCN        & ALECE          & NeuroCard     & FactorJoin & UAE         & DistJoin \\ \midrule
Training Time             & 24 h+351 s  & 24 h+1903 s    & 150+293 s     & 532+8 s    & 142s+5.6h   & 310 s    \\
Training GPU Mem.         & 2.47 GB     & 11.44 GB       & 4.49 GB       & 2.23 GB    & 5.49 GB     & 6.49 GB  \\
Inference Time            & 0.60 ms     & 82.76 ms       & 50.28 ms      & 8.97 ms    & 53.7 ms     & 26.56 ms \\
Inference GPU Mem.        & 2.27 GB     & 11.44 GB       & 4.34 GB       & 2.18 GB    & 4.99 GB     & 3.66 GB  \\
Total Models' Size        & 5.3 MB      & 61.72 MB       & 6.4 MB        & 2.3 MB     & 6.2 MB      & 56.1 MB  \\ \bottomrule
\end{tabular}
}
\end{table}

\subsection{Performance on Data Updating}
\begin{figure}[!htbp]
    \centering
    \includegraphics[width=\linewidth]{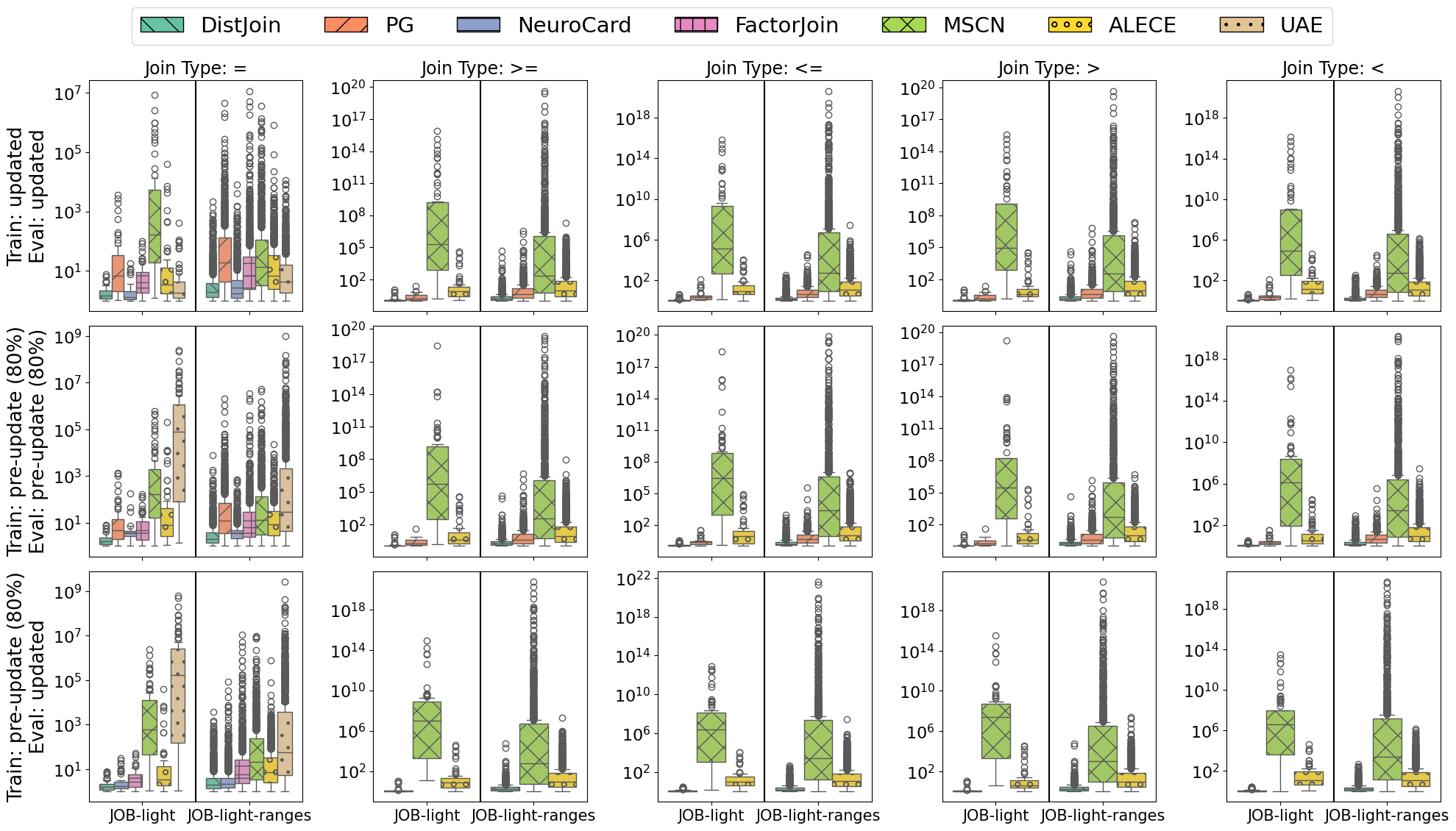}
    \caption{Q-Error comparison by workloads and join conditions to evaluate data update situation and robustness.}
    \vspace{-15pt}
    \label{fig:q_error_dataset}
\end{figure}

This subsection evaluates DistJoin's robustness against baselines. We simulate a pre-update dataset by randomly sampling 80\% of tuples from each table, train all methods on this dataset, and evaluate them on both pre-update and updated workloads. Results appear in \autoref{fig:q_error_dataset}.

For equi joins, DistJoin maintains the lowest Q-Error across all workload combinations. The last row of results, showing training on the pre-update dataset and evaluation on updated workloads, demonstrates DistJoin's superior robustness compared to baselines. For non-equi joins, DistJoin and PG show competitive performance on JOB-light under different join conditions, with each outperforming the other in specific scenarios. However, on JOB-light-ranges, DistJoin significantly outperforms PG. MSCN exhibits substantially higher errors across both workloads. 

The results show that, although DistJoin isn't designed for OLTP scenarios with high data updating speed, it still holds high accuracy during its periodic batch update, with an increase in median Q-Error of only 0.07 (1.99 vs 2.06) and in 99th Q-Error of only 44.79 (401.6 vs 446.39) on job-light-ranges, and outperforms all baseline estimators. While DistJoin's batch-oriented updates (232k tuples/s in experiments) cannot achieve millisecond-level latency, they suffice for OLTP scenarios where slightly stale estimates are acceptable.

\subsection{Ablation Study}
This section validates the ANPM model's effectiveness through an ablation study on workloads with equi join:

\textbf{(A)} Removing ANPM Blocks significantly increases Q-Error, proving ANPM's effectiveness in improving accuracy.

\textbf{(B)} Removing the \textit{SOS} and right-shift mechanism further increases Q-Error compared to (A).

\textbf{(C)} Removing the learned softmax temperature coefficient significantly increases JOB-light's maximum Q-Error while slightly decreasing other Q-Error metrics. Further investigation reveals that JOB-light includes equality predicates on highly skewed columns, and the filtered values' count is very small, demonstrating the learned temperature's effectiveness in handling skewed columns at the cost of slight accuracy reductions in other scenarios. Thus, enabling this feature selectively for skewed columns is optimal.

\textbf{(D)} Removing the mixed activation function improves performance across all scenarios. To further study the effect of it, we applied an extra ablation study \textbf{(E)}. 

\textbf{(E)} Removing only the mixed activation function from the base model to explore how its interaction with other features affects model accuracy. The result shows a slight improvement in JOB-light maximum Q-Error with significant decreases in other Q-Error metrics. This indicates that the mixed activation function should be used in conjunction with all three other features to enhance the model's performance in most scenarios.

\textbf{(F)} A hybrid approach using DistJoin's single-table probability estimation method combined with FactorJoin's cardinality inference approach (as shown in \autoref{equ:factor_equi_join}). The results demonstrate significant error growth compared to pure DistJoin, particularly exhibiting pronounced long-tail error distribution characteristics, which further validates the variance analysis conclusions presented in \autoref{section:error_analysis}.

\begin{table}[htbp]
\centering
\caption{Ablation study of DistJoin's model improvements on join condition $=$.}
\label{tab:ablation_study}
\resizebox{\linewidth}{!}{
\setlength{\tabcolsep}{3pt}
\begin{tabular}{cccccccccc}
\toprule
\multirow{2}{*}{\makecell{}} & \multirow{2}{*}{\makecell{ANPM\\Block}} & \multirow{2}{*}{\makecell{\textit{SOS}\\+Shift}} & \multirow{2}{*}{\makecell{Learned\\Temp.}}   & \multirow{2}{*}{\makecell{Mixed\\Act.}}   & \multirow{2}{*}{\makecell{Join\\Inference}}   & \multicolumn{2}{c}{JOB-light} & \multicolumn{2}{c}{JOB-light-ranges} \\ \cmidrule(lr){7-8} \cmidrule(lr){9-10}
        &                            &                                  &                 &                 &                               & 50th                  & max              & 50th             & max       \\ 
\midrule
base    &\ding{51}                   & \ding{51}                        & \ding{51}       & \ding{51}       & \multirow{5}{*}{DistJoin's}   & \textbf{1.49}         & 7.63             & \textbf{1.94}    & \textbf{2225.60}    \\
(A)     &                            & \ding{51}                        & \ding{51}       & \ding{51}       &                               & 1.69                  & 18.85            & 3.31             & $2.0e5$       \\
(B)     &                            &                                  & \ding{51}       & \ding{51}       &                               & 1.80                  & 19.09            & 3.37             & $2.6e5$    \\ 
(C)     &                            &                                  &                 & \ding{51}       &                               & 1.71                  & 128.10           & 3.14             & $1.4e5$     \\  
(D)     &                            &                                  &                 &                 &                               & 1.60                  & 125.90           & 2.87             & $1.5e5$      \\ \midrule
(E)     &\ding{51}                   & \ding{51}                        & \ding{51}       &                 & DistJoin's                    & 1.84                  & \textbf{6.97}    & 2.10             & 3709.30     \\  \midrule
(F)     &\ding{51}                   & \ding{51}                        & \ding{51}       & \ding{51}       & FactorJoin's                  & 2.58                  & 109.70           & 5.68             & $2.0e6$        \\

\bottomrule
\end{tabular}
}
\vspace{-15pt}
\end{table}

\section{Conclusion}
\label{section:Conclusion}
In this paper, we present Adaptive Neural Predicate Modulation (ANPM), a novel autoregressive model, and DistJoin, a decoupled join cardinality estimation framework built upon ANPM. DistJoin overcomes the ``Trilemma of Cardinality Estimation" by achieving state-of-the-art generality, accuracy, and updatability. Its \textit{selectivity-based join cardinality inference} method, enabled by ANPM's efficient probability estimation, extends support to non-equi joins-unprecedented in data-driven approaches. Theoretical and empirical analyses demonstrate that prior methods suffer from scalability constraints, whereas DistJoin's inference is provably unbiased (under given assumptions) and exhibits slower variance growth with increasing join size. Experiments confirm DistJoin's superiority in accuracy, robustness, and generality, alongside efficient inference and updatability. Ablation studies validate the enhancements introduced in ANPM. We are committed to further improving the joint density estimation efficiency of ANPM in future work, so that it can handle the approximate query processing scenario.

\section*{Acknowledgments}
This work is supported by National Natural Science Foundation of China (NSFC) (62232005).

During the preparation of this work the author(s) used DeepSeek~\cite{deepseekai2025deepseekv3technicalreport} in order to improve language readability and correct grammatical errors. After using this tool/service, the author(s) reviewed and edited the content as needed.

\vfill
\clearpage

\appendix
\section{Appendix} % 可以给总附录一个标题，或者直接在 \input 文件中开始具体附录章节
% 选择 \input 或 \include
\input{appendix.tex} % 推荐，因为它不会强制换页，更灵活

\end{document}

%% file: appendix.tex
%\documentclass[lettersize,journal]{IEEEtran}
% \usepackage{xr}
%\externaldocument{main}
%\usepackage{amsmath,amsfonts}
%\usepackage{algorithmic}
%\usepackage{algorithm}
%\usepackage{array}
% \usepackage[caption=false,font=normalsize,labelfont=sf,textfont=sf]{subfig}
%\usepackage{textcomp}
%\usepackage{stfloats}
%\usepackage{url}
%\usepackage{verbatim}
%\usepackage{graphicx}
%\usepackage{cite}
%\usepackage{amsmath}
%\usepackage{amsthm}
%\usepackage{pifont}
%\usepackage{diagbox}
%\usepackage{array}  
%\usepackage{multirow}
%\usepackage{booktabs}
%\usepackage{enumitem}
%\usepackage{tikz}
%\usepackage{subcaption}
%\usepackage{float}
%\usepackage{hyperref}
%\usepackage{xparse}
%\usepackage{mathtools}
%\usepackage[ruled, noend]{algorithm2e}
%\usepackage{makecell}
%\usepackage{siunitx} 
%\usepackage{etoolbox}
%\AfterEndEnvironment{algorithm}{\vspace{-1em}} 

% \SetAlgoSkip{SmallSkip}
%\setlength{\algomargin}{0em}
% updated with editorial comments 8/9/2021
%\newtheorem{assumption}{Assumption}
%\newtheorem{theorem}{Theorem}
%\newtheorem{lemma}[theorem]{Lemma}
%\newdefinition{rmk}{Remark}
%\newproof{pf}{Proof}
% \newproof{pot}{Proof of Theorem \ref{thm}}
%\NewDocumentEnvironment{pot}{m}{%
%  \noindent\textbf{Proof of Theorem~\ref{#1}.} 
%}{%
%  \qed
%}
%\newcommand{\assumptionautorefname}{Assumption}

\hyphenation{op-tical net-works semi-conduc-tor IEEE-Xplore}

%\begin{document}

\title{Appendix}

\maketitle

{\appendices
\section*{ANPM Block Formalized Calculation Process}
\label{section:ANPMBlock}

\begin{align}
    \overset{(B, F)}{x} &= \overset{(B, E)}{E(v_i^1)} \oplus \overset{(B, E)}{E(v_i^2)} \oplus \cdots \oplus \overset{(B, E)}{E(v_i^{j-1})} \label{equ:concat_training}\\
    \begin{split}
    \overset{(B, F)}{x} &= \Bigl(\dots\Bigl(\overset{(|\mathcal{D}(C_i^1)|, 1, E)}{E(v_i^1).U(1)} \oplus \overset{(1, |\mathcal{D}(C_i^1)|, E)}{E(v_i^2).U(0)}\Bigr).U(1) \\
    &\quad \oplus \cdots\Bigr) \oplus \overset{(1, |\mathcal{D}(1, C_i^1)|, E)}{E(v_i^{j-1}).U(0)}
    \end{split} 
    \label{equ:concat_inference}\\
\intertext{, where $F=(j-1)\times E$ and $B=\text{batch\_size}$ during training and $B=\prod_1^{j-1} |\mathcal{D}(C_i^j)|$ during inference. Then, the ANPM projects the logits by the following formulas:}
    \overset{(B, F, 1)}{u_h} &= U(\text{HyperNet}_h^u(x), 2) \\
    \overset{(B, 1, E)}{v_h} &= U(\text{HyperNet}_h^v(x), 1) \\
    \overset{(B, E)}{b_h} &= \text{HyperNet}_h^b(x) \\
    \overset{(B, E,1)}{u_l} &= U(\text{HyperNet}_l^u(x), 2) \\
    \overset{(B, 1, F)}{v_l} &=  U(\text{HyperNet}_l^v(x),1) \\ 
    \overset{(B, F)}{b_l} &= \text{HyperNet}_l^b(x) \\
    \overset{(B, F, E)}{W_h} &= u_h \times v_h, \quad \overset{(B, E, F)}{W_l} = u_l \times v_l \\
    \overset{(B, F)}{logits_i^{\prime j}} &= ReLU(\overset{(B,F)}{logits_i^j} \times W_h + b_h) \times W_l +b_l 
\end{align}
, where $U$ represents $unsqueeze$ function.

\section*{Proof of the Expectation and Variance of FactorJoin's Join Cardinality Inference Methods}
\label{section:ProofFactorJoin}
Before proving \autoref{thm:FactorJoinVariance}, we establish that neural network probability estimation errors follow an arbitrary distribution $\mathcal{G}$. This represents a relaxed constraint compared to typical assumptions of normal distribution in related works~\cite{DeepLearning, Adam, GELU}. Since our analysis focuses on join cardinality inference accuracy, we also assume unbiased probability estimates for individual tables, implying $\mathbb{E}[\mathcal{G}] = 0$.

\begin{assumption}
\label{assump:error_distribution}
For convenience and clarity, let the bin depth be 1, thus $bin(v)=v$ and $p_i(\cdot)$ is the true probability. We assume the error of the predicted probability follows:
\begin{align}
\hat{p}_{T_i}(T_i.C_{key}&=v,\mathcal{Q}(T_i))=p_i(v,\mathcal{Q})+\epsilon_i(v,\mathcal{Q})\\
\epsilon_i(v,\mathcal{Q}) &\sim \mathcal{G}(0,\sigma_i^2(v,\mathcal{Q})) \\ 
\hat{p}_{T_i}(T_i.C_{key}&=v)=p_i(v)+\epsilon_i(v)\\
\epsilon_i(v) &\sim \mathcal{G}(0,\sigma_i^2(v))
\end{align}
, where $\mathcal{G}(\mu,\sigma^2)$ represents an arbitrary distribution with expectation $\mu$ and variance $\sigma^2$.
\end{assumption}

With the \autoref{assump:error_distribution}, we proved the \autoref{thm:FactorJoinVariance} as follows:

\begin{pot}{thm:FactorJoinVariance}
Since each table has an independent estimator, the estimated probability of each table is independent. Thus , ~\autoref{equ:FactorJoinExp} can be proved by:
    \begin{align}
        \mathbb{E}[\hat{\text{card}}]
        &= \mathbb{E}\left[\sum_{v}\prod_{i=1}^{N}\hat{p}_{T_i}(T_i.C_{key}=v,\mathcal{Q}(T_i))|T_i|\right] \\
        &= \sum_{v}\prod_{i=1}^{N} |T_i|\mathbb{E}\left[\hat{p}_{T_i}(T_i.C_{key}=v,\mathcal{Q}(T_i))\right] \\ % && \text{by linearity of expectation}
        &=\sum_{v}\prod_{i=1}^{N} |T_i|p_i(v)
    \end{align}
    \par
    Next, we prove the variance is proportional to $\prod_{i=1}^n|T_i|^2$. Based on the designs that both FactorJoin and DistJoin employ independent models to estimate different tables, there are the following facts:
    \begin{enumerate}[label=\alph*)]
        \item Since the prediction errors among different key values of the same table are related: $\text{Cov}(\epsilon(v,\mathcal{Q}), \epsilon(v^\prime,\mathcal{Q}))=C_i^Q(v,v^\prime)$
        \item Since prediction errors among different tables are independent: $\text{Cov}(\epsilon_i(v,\mathcal{Q}), \epsilon_j(v^\prime,\mathcal{Q}))=0, \quad (i \neq j)$
    \end{enumerate}
    Expanding the ~\autoref{equ:factor_equi_join} by:
    \begin{align}
       \hat{\text{card}}
       &=\prod_{i=1}^n|T_i|\sum_v \prod_{i=1}^n\hat{p}_i(v,\mathcal{Q})\\
       &=\prod_{i=1}^n|T_i|\sum_v \prod_{i=1}^n\left( p_i(v,\mathcal{Q})+\epsilon_i(v,\mathcal{Q})\right)\\
       &=\prod_{i=1}^n|T_i| \cdot X \\
       X &= \sum_v \prod_{i=1}^n\left( p_i(v,\mathcal{Q})+\epsilon_i(v,\mathcal{Q})\right)\\
       \intertext{The variance of estimated cardinality is:}
       \text{Var}(\hat{\text{card}}) 
       &= \left(\prod_{i=1}^n|T_i|\right)^2\text{Var}\left(X\right)\label{equ:FactorJoinVariance}\\
       \intertext{By $X$ is independent of $|T_i|$ since $p_i$ is normalized probability}
       &\propto \left(\prod_{i=1}^n|T_i|\right)^2 
    \end{align}
\end{pot}

\section*{Proof of the Expectation and Variance of DistJoin's Join Cardinality Inference Methods}
\label{section:ProofDistJoin}
\begin{assumption}
\label{assump:DistJoinAnalysis}
For analytical convenience, we assume zero covariance between distribution estimates under different inputs:
\begin{equation}
    \text{Cov}(f(X_1),f(X_2)) = 0,\ \text{where}\ f(\cdot)\ \text{is an estimator}
\end{equation}
\end{assumption}

\begin{pot}{thm:DistJoinExpVar}
We first analyze the expectation and variance of the two-table joins situation:
\begin{align}
\hat{\text{card}} &= \text{card}_{J(\mathcal{T})} \frac{\sum_v \hat{p}_{T_l}(v,\mathcal{Q})\hat{p}_{T_r}(v,\mathcal{Q})}{\sum_v \hat{p}_{T_l}(v)\hat{p}_{T_r}(v)}\\
\intertext{By the ~\autoref{assump:error_distribution}:}
\mathbb{E}[\hat{p}_{T_l}(v,\mathcal{Q})] &= p_{T_l}(v,\mathcal{Q}), \quad \mathbb{E}[\hat{p}_{T_r}(v,\mathcal{Q})] = p_{T_r}(v,\mathcal{Q}) \\
\mathbb{E}[\hat{p}_{T_l}(v)] &= p_{T_l}(v), \quad \mathbb{E}[\hat{p}_{T_r}(v)] = p_{T_r}(v) \\
\intertext{Let:}
\hat{S_{\mathcal{Q}}} &= \sum_v \hat{p}_{T_l}(v,\mathcal{Q})\hat{p}_{T_r}(v,\mathcal{Q})\\
\hat{S} &= \sum_v \hat{p}_{T_l}(v)\hat{p}_{T_r}(v)\\
S_{\mathcal{Q}}& =\sum_v p_{T_l}(v,\mathcal{Q})p_{T_r}(v,\mathcal{Q})\\
S &= \sum_v p_{T_l}(v)p_{T_r}(v)\\
\Delta S_{\mathcal{Q}}&=\hat{S_{\mathcal{Q}}}-S_{\mathcal{Q}} ,\quad \Delta S=\hat{S}-S\\
\intertext{By using independent trained models for different tables and ~\autoref{assump:error_distribution}:}
\mathbb{E}[\hat{S_{\mathcal{Q}}}] &= \sum_v p_{T_l}(v,\mathcal{Q})p_{T_r}(v,\mathcal{Q}) \\
\mathbb{E}[\hat{S}] &= \sum_v p_{T_l}(v)p_{T_r}(v) \\
\intertext{Taylor expansion of $\frac{\hat{S_{\mathcal{Q}}}}{\hat{S}}$ at $\mathbb{E}[\hat{S}]$:}
\frac{\hat{S_{\mathcal{Q}}}}{\hat{S}}
&\approx \frac{\hat{S_{\mathcal{Q}}}}{\mathbb{E}[\hat{S}]} - \frac{\hat{S_{\mathcal{Q}}}(\hat{S}-\mathbb{E}[\hat{S}])}{\mathbb{E}[\hat{S}]^2} + \dots \\
\mathbb{E}\left[\frac{\hat{S_{\mathcal{Q}}}}{\hat{S}}\right]
&\approx \frac{\mathbb{E}[\hat{S_{\mathcal{Q}}}]}{\mathbb{E}[\hat{S}]} - \frac{\text{Cov}(\hat{S_{\mathcal{Q}}},\hat{S})}{\mathbb{E}[\hat{S}]^2} + \dots \label{equ:DistJoinTaylorExp}\\
\intertext{By ~\autoref{assump:DistJoinAnalysis}, independent estimation for different tables and ignore the higher-order terms}
&\approx \frac{\mathbb{E}[\hat{S_{\mathcal{Q}}}]}{\mathbb{E}[\hat{S}]} \label{equ:keyStepInvolveAssum} \\
&= \frac{\sum_v p_{T_l}(v,\mathcal{Q})p_{T_r}(v,\mathcal{Q})}{\sum_v p_{T_l}(v)p_{T_r}(v)}\\
\mathbb{E}[\hat{\text{card}}] 
&= \text{card}_{J(\mathcal{T})} \mathbb{E}[\frac{\hat{S_{\mathcal{Q}}}}{\hat{S}}] \\
&\approx \text{card}_{J(\mathcal{T})} \frac{\sum_v p_{T_l}(v,\mathcal{Q})p_{T_r}(v,\mathcal{Q})}{\sum_v p_{T_l}(v)p_{T_r}(v)} \\
&= \text{card}
\end{align}
% S_{\mathcal{Q}}& =\sum_v\prod p_i(v,\mathcal{Q}),\quad S=\sum_v\prod p_i(v)\\

%Since each step of the recursive inference can be regarded as the query's selectivity on a new joined table, and that result is approximately unbiased that subjects to ~\autoref{assump:error_distribution}. Thus, the result of expectation can be extended to the N-tables join situation. Now we analyze the variance under two-table joins situation:
\noindent By $\text{card}_{J(\mathcal{T})}$ is a fixed value that is calculated based on the true distribution, so it's independent of either $\hat{S_{\mathcal{Q}}}$ or $S$:
\begin{align}
\text{Var}(\hat{\text{card}})&=\text{Var}\left(\text{card}_{J(\mathcal{T})} \frac{\hat{S_{\mathcal{Q}}}}{\hat{S}}\right)
=\text{card}_{J(\mathcal{T})}^2 \text{Var}\left(\frac{\hat{S_{\mathcal{Q}}}}{\hat{S}}\right)\\
\intertext{By Taylor expansion and ignoring higher-order terms:}
\frac{\hat{S_{\mathcal{Q}}}}{\hat{S}}
&\approx \frac{S_{\mathcal{Q}}}{S}+ \frac{\Delta S_{\mathcal{Q}}}{S} - \frac{S_{\mathcal{Q}}}{S^2} \Delta S \label{equ:TaylorExp}\\
%&\approx 1+ \frac{\Delta \hat{S_{\mathcal{Q}}}}{\hat{S}}\\
\begin{split}
\text{Var}(\hat{\text{card}})
&= \text{card}_{J(\mathcal{T})}^2 \text{Var}\left(\frac{\hat{S_{\mathcal{Q}}}}{\hat{S}}\right)\\
&\approx \text{card}_{J(\mathcal{T})}^2 \left[\frac{\text{Var}(\Delta S_\mathcal{Q})}{S^2} + \frac{S_{\mathcal{Q}}^2}{S^4}\text{Var}(\Delta S) \right. \\
&\quad \left. -2\frac{S_{\mathcal{Q}}}{S^3}\text{Cov}(\Delta S_{\mathcal{Q}}, \Delta S)\right]\label{equ:DistJoinVariance} \\
&\propto\text{card}_{J(\mathcal{T})}^2
\end{split}
\end{align}
%Although $\text{Var}\left(\frac{\hat{S_{\mathcal{Q}}}}{\hat{S}}\right)$ may increase as different number of joins, this result can still be naturally extended to the N-tables join situation since $\text{Var}\left(\frac{\hat{S_{\mathcal{Q}}}}{\hat{S}}\right)$ is independent of $\text{card}_{J(\mathcal{T})}^2$.

We can use induction to extend the conclusions above to n-table joins. For $k+1$ table joins situation: $T_1 \Join \dots \Join T_{k+1}$, we regard it as $\mathcal{T}^{(k)} \Join T_{k+1}$, the cardinality can be calculated by:
\begin{align} \label{equ:RecursionCard}
    \begin{split}
        \hat{\text{card}}^{(k+1)} &= \text{card}_{J(\mathcal{T}^{(k+1)})} \cdot \\
        &\quad \frac{
            \sum_v \hat{p}_{\mathcal{T}^{(k)}}\left(v,\mathcal{Q}\left(\mathcal{T}^{(k)}\right)\right) \cdot \hat{p}_{T_{k+1}}\left(v,\mathcal{Q}\left(T_{k+1}\right)\right)
        }{
            \sum_v \hat{p}_{\mathcal{T}^{(k)}}(v) \hat{p}_{T_{k+1}}(v)
        }\\
    \end{split}
\end{align}
\begin{align}
    \text{card}_{J(\mathcal{T}^{(k+1)})} &= \sum_v \left[\text{card}_{J(\mathcal{T}^{(k)})} \cdot p_{\mathcal{T}^{(k)}}(v)\right] \cdot \left[|T_{k+1}| \cdot p_{T_{k+1}}(v)\right]   
\end{align}

Since $\text{card}_{J(\mathcal{T}^{(k+1)})}$ is a fixed true value, we only need to prove that $\hat{p}_{\mathcal{T}^{(k)}}$ follows ~\autoref{assump:error_distribution}. We proceed by mathematical induction:

\noindent \textbf{Base Case ($ k = 1 $):}
Based on ~\autoref{assump:error_distribution}:
\begin{align}
    \hat{p}_{\mathcal{T}^{(1)}}(v,\mathcal{Q}) &= p_{\mathcal{T}^{(1)}}(v,\mathcal{Q}) + \epsilon_{\mathcal{T}^{(1)}}(v,\mathcal{Q}) \\
    \epsilon_{\mathcal{T}^{(1)}}(v,\mathcal{Q}) &\sim \mathcal{G}^{(1)}(0,\sigma_1^2(v,\mathcal{Q}))\\
    \hat{p}_{\mathcal{T}^{(1)}}(v) &= p_{\mathcal{T}^{(1)}}(v) + \epsilon_{\mathcal{T}^{(1)}}(v) \\
    \epsilon_{\mathcal{T}^{(1)}}(v) &\sim \mathcal{G}^{(1)}(0,\sigma_1^2(v))
\end{align}
\noindent \textbf{Inductive Hypothesis:}
Assuming the $k-1$-th join holds:
\begin{align}
    \hat{p}_{\mathcal{T}^{(k-1)}}(v,\mathcal{Q})&=p_{\mathcal{T}^{(k-1)}}(v,\mathcal{Q}) + \epsilon_{\mathcal{T}^{(k-1)}}(v,\mathcal{Q}) \\
    \epsilon_{\mathcal{T}^{(k-1)}}(v,\mathcal{Q}) &\sim \mathcal{G}^{(k-1)}(0,\sigma_{k-1}^2(v,\mathcal{Q}))\\
    \hat{p}_{\mathcal{T}^{(k-1)}}(v)&=p_{\mathcal{T}^{(k-1)}}(v) + \epsilon_{\mathcal{T}^{(k-1)}}(v) \\
    \epsilon_{\mathcal{T}^{(k-1)}}(v) &\sim \mathcal{G}^{(k-1)}(0,\sigma_{k-1}^2(v))
\end{align}
\noindent \textbf{Inductive Step:}
We need to prove that for the $k$-th join, it holds:
\begin{align}
    \hat{p}_{\mathcal{T}^{(k)}}(v,\mathcal{Q})&=p_{\mathcal{T}^{(k)}}(v,\mathcal{Q}) + \epsilon_{\mathcal{T}^{(k)}}(v,\mathcal{Q}) \\
    \epsilon_{\mathcal{T}^{(k)}}(v,\mathcal{Q}) &\sim \mathcal{G}^{(k)}(0,\sigma_{k}^2(v,\mathcal{Q}))\\
    \hat{p}_{\mathcal{T}^{(k)}}(v)&=p_{\mathcal{T}^{(k)}}(v) + \epsilon_{\mathcal{T}^{(k)}}(v)\\
    \epsilon_{\mathcal{T}^{(k)}}(v) &\sim \mathcal{G}^{(k)}(0,\sigma_{k}^2(v))
\end{align}
Since in ~\autoref{assump:error_distribution}, $\mathcal{G}^{(k)}$ is an arbitrary distribution and $\sigma_{k}^2$ is an arbitrary variance, we only need to prove that the expectation of the error $\epsilon_{\mathcal{T}^{(k)}}$ is zero.\\
By ~\autoref{equ:multi_tables_equi_join}, we have:
\begin{align}
    \hat{p}_{\mathcal{T}^{(k)}}(v,\mathcal{Q}) &= \frac{\hat{p}_{\mathcal{T}^{(k-1)}}\left(v,\mathcal{Q}\left(\mathcal{T}^{(k-1)}\right)\right) \cdot  \hat{p}_{T_{k}}\left(v,\mathcal{Q}\left(T_{k}\right)\right)}{\sum_v \hat{p}_{\mathcal{T}^{(k-1)}}(v) \hat{p}_{T_{k}}(v) }
\end{align}
Let:
\begin{align}
    \hat{S}_Q^{(k)}&=\hat{p}_{\mathcal{T}^{(k-1)}}\left(v,\mathcal{Q}\left(\mathcal{T}^{(k-1)}\right)\right) \cdot  \hat{p}_{T_{k}}\left(v,\mathcal{Q}\left(T_{k}\right)\right) \\
    \hat{S}^{(k)} &=\sum_v \hat{p}_{\mathcal{T}^{(k-1)}}(v) \hat{p}_{T_{k}}(v)\\
    S_{\mathcal{Q}}^{(k)}&=p_{\mathcal{T}^{(k-1)}}\left(v,\mathcal{Q}\left(\mathcal{T}^{(k-1)}\right)\right) \cdot p_{T_{k}}\left(v,\mathcal{Q}\left(T_{k}\right)\right) \\
    S^{(k)} &=\sum_v p_{\mathcal{T}^{(k-1)}}(v) p_{T_{k}}(v)\\
    \Delta S_{\mathcal{Q}}^{(k)} &= \hat{S}_Q^{(k)} - S_{\mathcal{Q}}^{(k)}, \qquad \Delta S^{(k)} = \hat{S}^{(k)} - S^{(k)}
\end{align}
Similar to the formula~\autoref{equ:DistJoinTaylorExp}, by Taylor expansion, we can be obtained:
\begin{align}
    \mathbb{E}\left[\hat{p}_{\mathcal{T}^{(k)}}(v,\mathcal{Q})\right]
    &\approx \frac{\mathbb{E}\left[\hat{S_{\mathcal{Q}}}^{(k)}\right]}{\mathbb{E}\left[\hat{S}^{(k)}\right]} - \frac{\text{Cov}\left(\hat{S_{\mathcal{Q}}}^{(k)},\hat{S}^{(k)}\right)}{\mathbb{E}\left[\hat{S}^{(k)}\right]^2} + \dots\\
\intertext{By ~\autoref{assump:DistJoinAnalysis} and independent estimation between different tables:}
    &\approx \frac{\mathbb{E}\left[\hat{S_{\mathcal{Q}}}^{(k)}\right]}{\mathbb{E}\left[\hat{S}^{(k)}\right]}
\intertext{By using independent trained models for different tables and the inductive hypothesis:}
    &=\frac{S_{\mathcal{Q}}^{(k)}}{S^{(k)}}\\
\intertext{Because :}
    \mathbb{E}\left[\hat{p}_{\mathcal{T}^{(k)}}(v,\mathcal{Q})\right]
    &=\mathbb{E}\left[p_{\mathcal{T}^{(k)}}(v,\mathcal{Q}) + \epsilon_{\mathcal{T}^{(k)}}(v,\mathcal{Q})\right]\\
    &=p_{\mathcal{T}^{(k)}}(v,\mathcal{Q}) + \mathbb{E}\left[\epsilon_{\mathcal{T}^{(k)}}(v,\mathcal{Q})\right]\\
    &=\frac{S_{\mathcal{Q}}^{(k)}}{S^{(k)}} + \mathbb{E}\left[\epsilon_{\mathcal{T}^{(k)}}(v,\mathcal{Q})\right]\\
\intertext{We can obtain that:}
    \mathbb{E}\left[\epsilon_{\mathcal{T}^{(k)}}(v,\mathcal{Q})\right] &= 0
\end{align}
\textbf{End of Induction Proof.} Thus, because the base case holds and the inductive step holds, we proved that for $N$-table joins, the $\epsilon_{\mathcal{T}^{(k)}}(v,\mathcal{Q}) \sim \mathcal{G}^{(k)}(0,\sigma_{k}^2(v,\mathcal{Q}))$ holds. By the same reasoning, let $\mathcal{Q}=\emptyset$, the $\epsilon_{\mathcal{T}^{(k)}}(v) \sim \mathcal{G}^{(k)}(0,\sigma_{k}^2(v))$ holds. That is, $\hat{p}_{\mathcal{T}^{(k)}}\left(v,\mathcal{Q}\left(\mathcal{T}^{(k)}\right)\right)$ follows the ~\autoref{assump:error_distribution}.

\noindent \textbf{Now we use induction to extend the conclusions of two-table joins to N-table joins:\\}
\noindent \textbf{Base Case ($ k = 2 $):}
\begin{align}
    \mathbb{E}[\hat{\text{card}}^{(2)}] &\approx \text{card}_{J(\mathcal{T}^{(2)})} \frac{\sum_v p_{T_l}(v,\mathcal{Q})p_{T_r}(v,\mathcal{Q})}{\sum_v p_{T_l}(v)p_{T_r}(v)} \\
    &= \text{card}^{(2)}\\
            \text{Var}(\hat{\text{card}}^{(2)}) &\propto \text{card}_{J(\mathcal{T}^{(2)})}^2    
\end{align}
\noindent \textbf{Inductive Hypothesis:}
Assuming the $k$-table joins hold:
\begin{align}
    \mathbb{E}[\hat{\text{card}}^{(k)}] &\approx \text{card}_{J(\mathcal{T}^{(k)})} \frac{\sum_v p_{\mathcal{T}^{(k-1)}}(v,\mathcal{Q})p_{T_k}(v,\mathcal{Q})}{\sum_v p_{\mathcal{T}^{(k-1)}}(v)p_{T_k}(v)} \\
    &= \text{card}^{(k)}\\
    \text{Var}(\hat{\text{card}}^{(k)}) &\propto \text{card}_{J(\mathcal{T}^{(k)})}^2   
\end{align}
\noindent \textbf{Inductive Step:}
Based on ~\autoref{equ:RecursionCard}, ~\autoref{assump:DistJoinAnalysis}, and the conclusion we just proved that $\hat{p}_{\mathcal{T}^{(k)}}\left(v,\mathcal{Q}\left(\mathcal{T}^{(k)}\right)\right)$ follows the ~\autoref{assump:error_distribution}, we can regard $\mathcal{T}^{(k-1)}$ as an individual table. Then, with the same reason that we proved the base case, the following formulas hold for $n=k$ situation.
\begin{align}
    \mathbb{E}[\hat{\text{card}}^{(k)}] &\approx \text{card}_{J(\mathcal{T}^{(k)})} \frac{\sum_v p_{\mathcal{T}^{(k-1)}}(v,\mathcal{Q})p_{T_{k}}(v,\mathcal{Q})}{\sum_v p_{\mathcal{T}^{(k-1)}}(v)p_{T_{k}}(v)} \\
    &= \text{card}^{(k)}\\
    \text{Var}(\hat{\text{card}}^{(k)}) &\propto \text{card}_{J(\mathcal{T}^{(k)})}^2   
\end{align}
\textbf{End of Proof.} In summary, we first proved that ~\autoref{thm:DistJoinExpVar} holds for the two-table joins situation, and then we used induction and proved that the conclusion can be extended to the n-table joins situation. Thus, we have proven ~\autoref{thm:DistJoinExpVar}.
\end{pot}

\section*{Algorithm of how DistJoin deals with non-equi joins}
\label{section:DistJoinForNonEquiJoins}
\begin{algorithm}[!htbp]
    \caption{Inference of Join Cardinality for Different Join Conditions}
    \label{alg:InferenceForJoinTypes}
    \LinesNumbered
    \footnotesize
    \KwIn{probability distributions: $\hat{\mathbf{P}}$}
    %$\hat{\mathbf{P}}_{\mathcal{T}_{left}}(C_{key}, \mathcal{Q}), \hat{\mathbf{P}}_{\mathcal{T}_{right}}(C_{key}, \mathcal{Q}),\hat{\mathbf{P}}_{\mathcal{T}_{left}}(C_{key}),\hat{\mathbf{P}}_{\mathcal{T}_{right}}(C_{key})$}
    \KwOut{$\hat{\mathbf{P}}_{\mathcal{T}}(C_{key}, \mathcal{Q})$}
    $p_l(\mathcal{Q}),p_r(\mathcal{Q}),p_l,p_r \longleftarrow \hat{\mathbf{P}}_{\mathcal{T}_{left}}(C_{key}, \mathcal{Q}), \hat{\mathbf{P}}_{\mathcal{T}_{right}}(C_{key}, \mathcal{Q}),\hat{\mathbf{P}}_{\mathcal{T}_{left}}(C_{key}),\hat{\mathbf{P}}_{\mathcal{T}_{right}}(C_{key})$; \\
    \uIf{$join\_how='\geq'$}{
        $p_l(\mathcal{Q}) \longleftarrow p_l(\mathcal{Q}) + \sum p_l(\mathcal{Q}) - \text{cumsum}(p_l(\mathcal{Q}))$ ; \text{//reverse cumsum}\\
    }
    \uIf{$join\_how='\leq'$}{
        $p_l(\mathcal{Q}) \longleftarrow \text{cumsum}(p_l(\mathcal{Q}))$ ;\\
    }
    \uIf{$join\_how='>'$}{
        $p_l(\mathcal{Q}) \longleftarrow \text{Roll}(p_l(\mathcal{Q}), -1)$ ;  \text{//Circularly shift left one position}\\
        $p_l(\mathcal{Q})[-1] \longleftarrow 0;$\\
        $p_l(\mathcal{Q}) \longleftarrow p_l(\mathcal{Q}) + \sum p_l(\mathcal{Q}) - \text{cumsum}(p_l(\mathcal{Q}))$ ; \text{//reverse cumsum}\\
    }
    \uIf{$join\_how='<'$}{
        $p_l(\mathcal{Q}) \longleftarrow \text{Roll}(p_l(\mathcal{Q}), 1)$ ;  \text{//Circularly shift right one position}\\
        $p_l(\mathcal{Q})[0] \longleftarrow 0;$\\
        $p_l(\mathcal{Q}) \longleftarrow \text{cumsum}(p_l(\mathcal{Q}))$ ;\\
    }
    $\hat{\mathbf{P}}_{\mathcal{T}}(C_{key}, \mathcal{Q}) \longleftarrow \frac{p_l(\mathcal{Q}) \odot p_r(\mathcal{Q})}{\sum_v p_l \cdot p_r}$ ;\\
    
    \Return $\hat{\mathbf{P}}_{\mathcal{T}}(C_{key}, \mathcal{Q})$;\\
\end{algorithm}

\section*{Additional Experiment}

\subsection*{Supplementary Settings}

\textbf{\underline{Model Architectures.}} Each table's ANPM model includes:
\begin{itemize}
    \item ResMADE with 4 layers (256 units each)
    \item ANPM HyperNetworks with 2 layers
    \item Factorized value bit width: 12
    \item Predicate embedding size: 64
\end{itemize}

Since the ANPM Blocks are the performance bottleneck, we also list the parameter numbers of each model's ANPM Blocks in \autoref{tab:ANPM_parameters}. The results demonstrate that despite the large \#DV values for each table, ANPM's low-rank architecture ensures the corresponding model maintains a remarkably small number of ANPM parameters.

\begin{table}[!htbp]
    \centering
    \caption{Number of ANPM Blocks' parameters.}
    \label{tab:ANPM_parameters}
    \begin{tabular}{cccccc}
        \toprule
        Model of table               & Parameter Numbers  \\
        \midrule
        \textit{cast\_info}          & 49920              \\
        \textit{movie\_companies}    & 99840              \\  
        \textit{movie\_info}         & 49920              \\
        \textit{movie\_info\_idx}    & 49920              \\
        \textit{movie\_keyword}      & 99840              \\
        \textit{title}               & 127140              \\
        \bottomrule
    \end{tabular}
\end{table}

\subsection*{Workload Distribution}
The workloads' distributions are shown as \autoref{fig:workload_dist}
\begin{figure*}[!htbp]
    \centering
    \includegraphics[width=\linewidth]{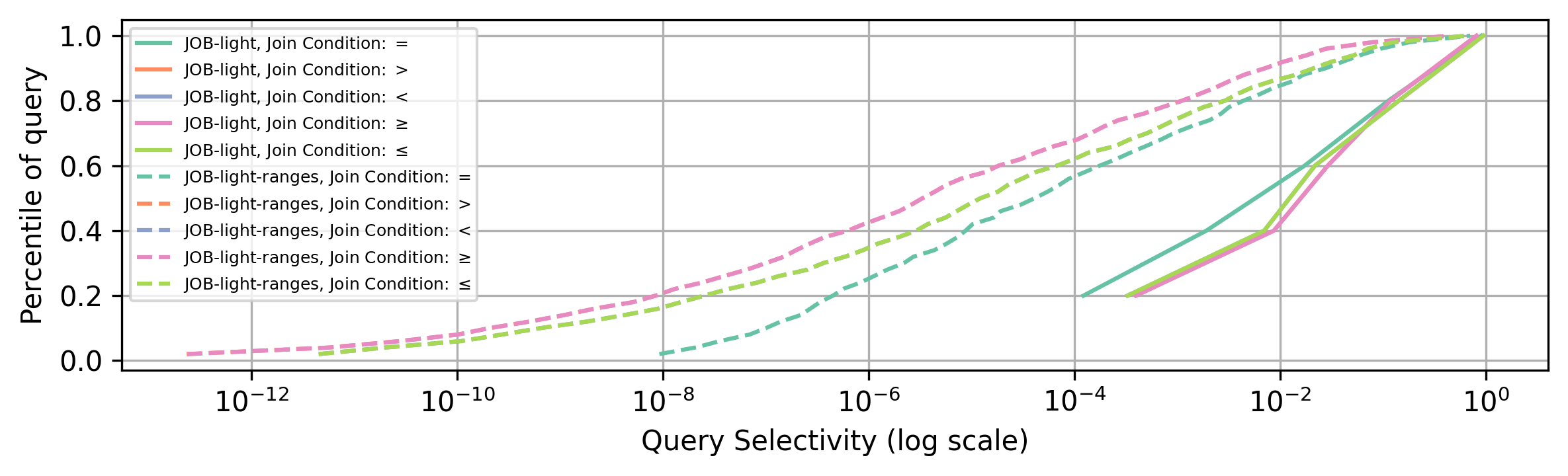}
    \caption{Distribution of each workload. The join conditions $(>,\geq)$ and $(<,\leq)$ have very similar selectivity distribution, so their curves overlay each other.}
    \label{fig:workload_dist}
\end{figure*}

\subsection*{Comprehensive Q-Error comparison}
\label{section:fullcomparison}
We demonstrate the comprehensive Q-Error comparison of all join types and more detailed percentiles in \autoref{fig:q_error_dataset_full}. The results show that DistJoin outperforms all baselines on almost all Q-Error percentiles.

\begin{figure*}[!htbp]
    \centering
    \includegraphics[width=\linewidth]{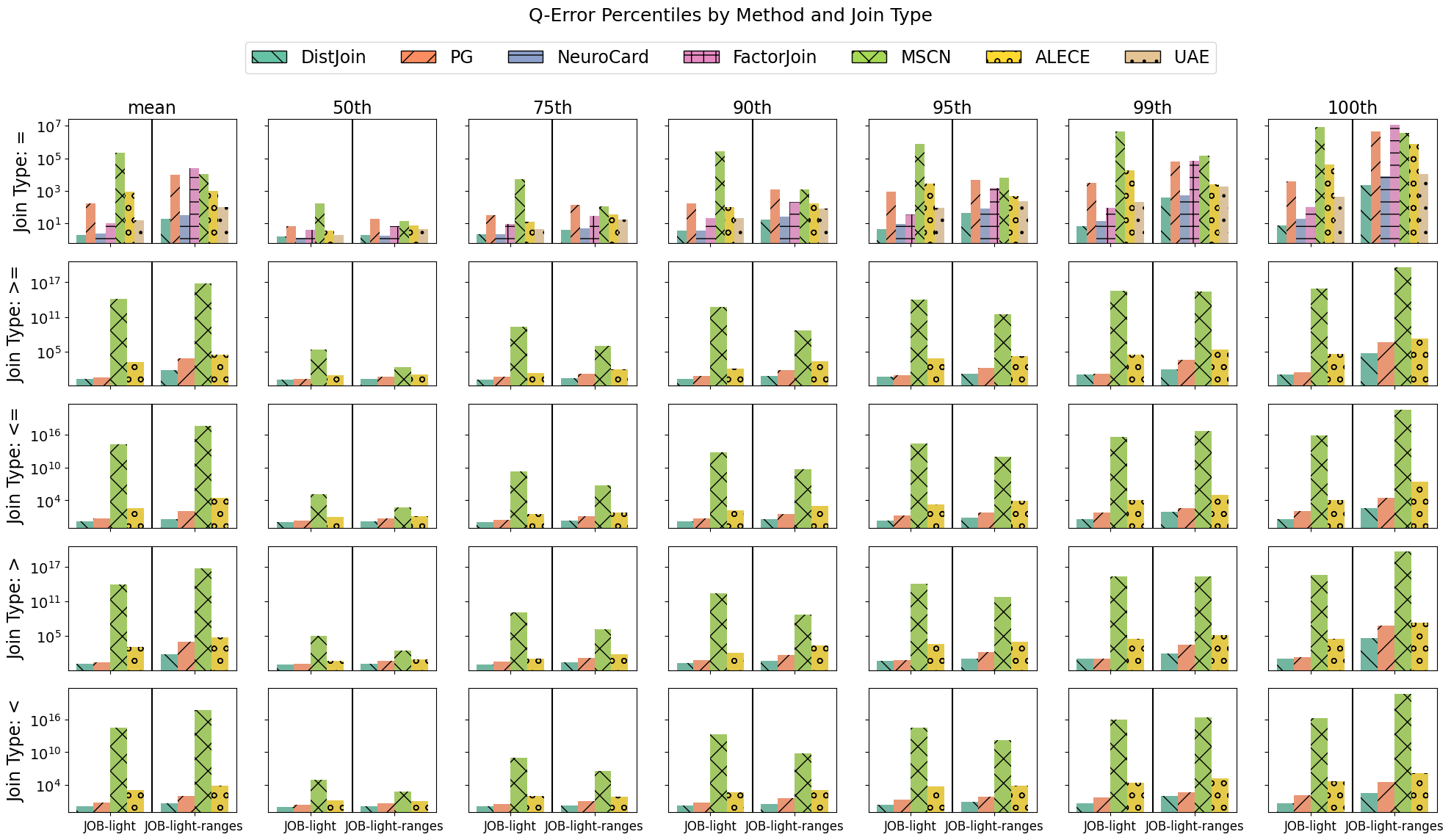}
    \caption{Q-Error comparison by workloads and join conditions to evaluate data update situation and robustness.}
    \label{fig:q_error_dataset_full}
\end{figure*}

\subsection*{Comparison on TPC-H datasets}
We generated complex join queries with range predicates on the TPC-H dataset to compare the most representative baseline's accuracy with DistJoin's. Following Naru's workload generation method~\cite{Naru}, we randomly selected join tables and then generated predicates on each of them. Finally, we calculate the true cardinality of them based on the equi join condition. The results in ~\autoref{tab:comparisonTPCH} show that DistJoin also significantly outperforms those baselines.
%DIFADDCMD <
\begin{table}[!htbp]
    \centering
    \caption{Q-Error comparison on TPC-H}
    \label{tab:comparisonTPCH}
    \begin{tabular}{cccccc}
        \toprule
        Methods                     & mean            & 50th          & 95th    & 99th     & max     \\ \midrule
        PG                          & 2.47            & \textbf{1.33} & 6.22    & 17.01    & 82.67   \\
        FactorJoin                  & $6.7e11$        & 8.57          & 211.26  & $4.0e8$  & $6.6e14$  \\
        DistJoin                    & \textbf{2.19}   & 1.50          & \textbf{5.67}      & \textbf{12.50}   & \textbf{38.0} \\
        \bottomrule
    \end{tabular}
\end{table}
%DIFADDCMD >

\subsection*{Evaluation of Convergence Speed}
We illustrate DistJoin's convergence curve in \autoref{fig:distjoin_convergence}. Since all tables train in parallel, grouping models by epoch proves difficult. Instead, we use the longest training time among all tables per epoch as the x-axis to demonstrate DistJoin's performance convergence over 20 epochs. DistJoin converges quickly in terms of median metrics, suggesting that for scenarios less sensitive to long-tail error distributions, training time can be reduced by accepting slightly increased long-tail error distributions for faster training.

\begin{figure*}[!htbp]
    \centering
    \includegraphics[width=\linewidth]{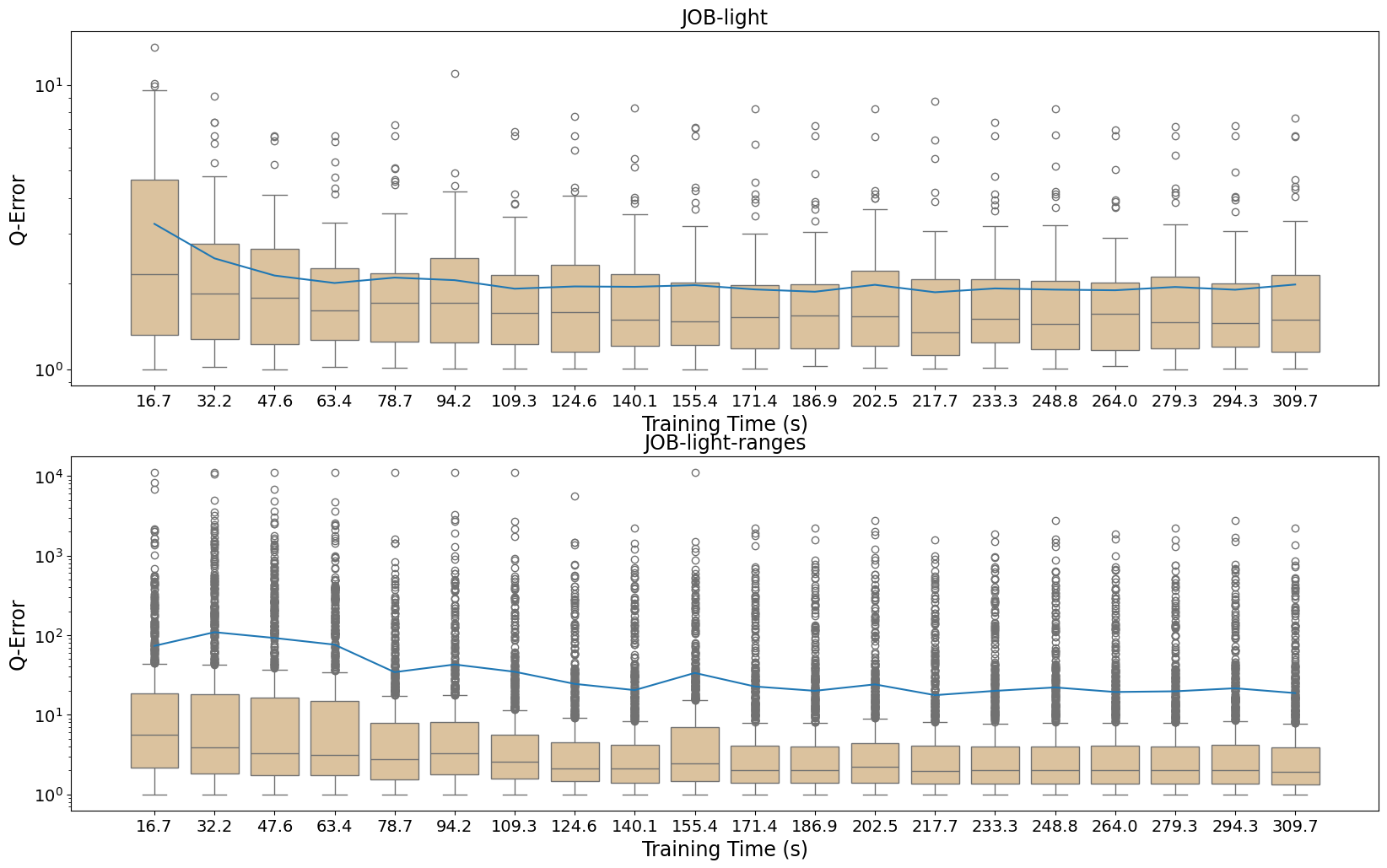}
    \caption{DistJoin's convergence curve.}
    \label{fig:distjoin_convergence}
\end{figure*}

}

\vfill

%\end{document}